\documentclass[jnr]{iosart2x}
\makeatletter
\let\numberlines@hook\relax
\makeatother

\usepackage{graphicx}
\usepackage{dcolumn}
\usepackage{bm}
\usepackage{hyperref}
\usepackage[utf8]{inputenc}
\usepackage[T1]{fontenc}
\usepackage{enumerate}
\usepackage{color}

\pubyear{2019}
\volume{0}
\firstpage{1}
\lastpage{1}
\let\originalleft\left
\let\originalright\right
\renewcommand{\left}{\mathopen{}\mathclose\bgroup\originalleft}
\renewcommand{\right}{\aftergroup\egroup\originalright}

\let\oldsqrt\sqrt
\def\sqrt{\mathpalette\DHLhksqrt}
\def\DHLhksqrt#1#2{%
\setbox0=\hbox{$#1\oldsqrt{#2\,}$}\dimen0=\ht0
\advance\dimen0-0.3\ht0
\setbox2=\hbox{\vrule height\ht0 depth -\dimen0}%
{\box0\lower0.4pt\box2}}

\newcommand{\parh}[1]{\left( #1 \right) }
\newcommand{\sbrac}[1]{\left[ #1 \right] }

\renewcommand{\ln}[1]{\mathrm{ln}\parh{#1}}

\newcommand{\pdx}[2][]{\frac{\partial #1}{\partial #2}}

\begin{document}

\begin{frontmatter}

\title{Multinomial, Poisson and Gaussian statistics in count data analysis}


\author[A,B]{\inits{K.}\fnms{Jakob} \snm{Lass}\ead[label=e1]{jakob.lass@nbi.ku.dk}},
\author[A]{\inits{M. E.}\fnms{Magnus Egede} \snm{B\o ggild}\ead[label=e1]{meb@nbi.ku.dk}},
\author[A]{\inits{P.}\fnms{Per} \snm{Hedegård}\ead[label=e1]{hedegard@nbi.ku.dk}},
and
\author[A]{\inits{K.}\fnms{Kim} \snm{Lefmann}\ead[label=e2]{lefmann@nbi.ku.dk}\thanks{Corresponding author. }}
\runningauthor{J. Lass et al.}

\address[A]{Niels Bohr Institute, \orgname{University of Copenhagen},
\cny{Denmark}\printead[presep={\\}]{e2}}
\address[B]{Laboratory for Neutron Scattering \orgname{Paul Scherrer Institute}
\cny{Switzerland}}
\begin{abstract}
It is generally known that counting statistics is not correctly described by a Gaussian approximation.
Nevertheless, in neutron scattering, it is common practice to apply this approximation to the counting statistics; also at low counting numbers. We show that the application of this approximation leads to skewed results not only for low-count features, such as background level estimation, but also for its estimation at double-digit count numbers. In effect, this approximation is shown to be imprecise on all levels of count. Instead, a Multinomial approach is introduced as well as a more standard Poisson method, which we compare with the Gaussian case. These two methods originate from a proper analysis of a multi-detector setup and a standard triple axis instrument.


We devise a simple mathematical procedure to produce unbiased fits using the Multinomial distribution and demonstrate this method on synthetic and actual inelastic scattering data. We find that the Multinomial method provide almost unbiased results, and in some cases outperforms the Poisson statistics. Although significantly biased, the Gaussian approach is in general more robust in cases where the fitted model is not a true representation of reality. For this reason, a proper data analysis toolbox for low-count neutron scattering should therefore contain more than one model for counting statistics.
\end{abstract}

\begin{keyword}
\kwd{Poisson statistics}
\kwd{Multinomial statistics}
\kwd{Data analysis}
\kwd{Neutron scattering}
\end{keyword}

\end{frontmatter}

\date{\today}

\maketitle


\section{Introduction}

The nature of the physical sciences is to apply a hypothesis to a system, such that it is possible to either confirm its accuracy, or falsify it, based on observation \cite{Barlow}. Usually, this observation consists of physically measured data which necessitates a statistical analysis, the type of which depends on the observation in question. In this article, we investigate analysis methods for low-statistics counting measurements, in particular inelastic neutron scattering data. Here, the current common practice is, due to convenience, to utilize the Gaussian limit of the Poisson statistics. This limit allows for the evaluation of fits by using the least squares method for which many algorithms are radially available, and to enable easier data transformation and normalisation. The approximative nature of the Gaussian treatment is well known and some software libraries are equipped to perform both the least squares method as well as the statistically correct Poisson treatment, e.g. MANTID \cite{Mantid}.

Numerous previous studies of counting statistics and their influence on Poisson parameter estimation have been published both in the statistical case, see e.g. Ref.~\cite{Patil2012}, or in the case of both single crystal and powder diffraction \cite{Hill1984}. In the latter case, both the low and high count limits are of concern, with the high limit being more common in the elastic case. The low limit results in wrong estimation of the counting uncertainty when using the fitting method of Gaussian least squares. However, in the high count regime, the counting uncertainty no longer provides the main source of error and thus, counting ''too`` long results in an underestimation of the uncertainties. This, in turn, obscures and possibly falsifies the parameter uncertainty in the presence of systematic errors originating from the experimental setup, an oversimplification in the model utilized, or other sources \cite{Hill1984}. We will here only be concerned with the question of statistical uncertainty, which will interchangingly be denoted as uncertainty and error.

In this article, we deal with the low-count limit of the Gaussian approximation, which we denote the {\em Poisson regime}. This is usually taken to be the regime with $N=10$ or less counts \cite{Barlow}. However, we show that the inaccuracies in the Gaussian parameter fitting in fact extend well outside this Poisson regime, their relative systematic error in the case of a constant background diminishing only as $1/N$. We discuss the merit of using alternative true Poisson and multinomial fitting methods and pinpoint the advantages and drawbacks of all methods.





\section{Model fitting by Gaussian and Poisson statistics}
Parameter estimation of a suggested model given a data set can be seen as a problem particularly well suited for the Bayesian approach. Using this method, it is possible to update the estimates of model parameters, given particular observations. That is, given the initial, or {\em a priory}, information $I$ for a model and set of parameters $M$, one updates their probabilities given the measured data $D$, according to Bayes Theorem \cite{Barlow}
\begin{equation}
P(M|DI)=\frac{P(D|MI)}{P(D|I)}P(M|I). \label{eq:BayesTheorem}
\end{equation}
Here $P(M|I)$ represents the initial probabilities of the model and its parameters, $P(D|I)$ is the probability of obtaining the observed data, $P(D|MI)$ gives the probability of the observation assuming a specific model and parameters, and finally $P(M|DI)$ is the posterior estimation of probabilities for the model and parameters. In order to apply this formula in practice, the probability of obtaining the data given the model needs to be found. As the model parameters are the intended result of the experiment, one needs to perform a fit that obtains these. This can be done in the Bayesian formalism by updating the parameter estimates with the new data as described in eq.~\eqref{eq:BayesTheorem}. However, in the case where no prior parameter values are more likely than others, one models this with a top hat prior. This requires the parameter to be finite, and as this usually is the case the prior can be set to be flat within the range of sensible values. This, in turn, makes the term $P(M|I)$ constant for all plausible parameter values. As $P(D|I)$ can be seen as a normalization constant independent of the model, we have
\begin{equation}
    P(D|MI) \propto P(M|DI).
\end{equation}
To optimize the probability, one simply optimizes the so-called {\em likelihood} term, $P(M|DI)$. This is in practice done by minimizing the negative log-likelihood:
\begin{align}
-\ln{P(M|D)}=-\ln{P(D|M)}&+\ln{P(M)}-\ln{P(D)}.
\end{align}
For both brevity and clarity $I$ has been removed from the above equation, as the symbol will later be used to denote the measured neutron count.

\subsection{The Poisson and Gaussian distributions}

Because of the discrete and uncorrelated nature of counting statistics, it is known that it follows the Poisson distribution \cite{Barlow}
with the probability of observing $n$ counts for a process that has an expected mean count of $\lambda$,
\begin{equation}
    P_P(n|\lambda) = \frac{\lambda^ne^{-\lambda}}{n!},
    \label{eq:PoissonDistribution}
\end{equation}
with a standard deviation given by $\sqrt{\lambda}$ \cite{Barlow}.
In the case of large mean counts, the Poisson distribution tends towards a Gaussian distribution, which also has mean $\lambda$ and standard deviation $\sigma=\sqrt{\lambda}$, i.e.
\begin{equation}
   P_P(n=x|\lambda) \approx \frac{1}{\sqrt{2 \pi \lambda}}\mathrm{e}^{-\frac{(x-\lambda)^2}{2\lambda}}.
\end{equation}

\subsection{Statistics on Scattering}
For simplicity, let us limit our discussion to reactor-based instruments with a monochromatic incoming beam. In most triple axis instruments, the process of measuring the scattering intensities, or more correctly the scattering cross section, for different processes in a material is either performed through a series of scans or with a multi-detector setup. Multi-detectors are also used for SANS, imaging, and powder diffraction. Here each detector (or detector pixel) corresponds to a specific momentum transfer $Q$ (and possibly energy transfer $E$). Despite the apparent differences the resulting statistics is the same. This can be seen by first considering the case of a point by point measurement. At each setting, one of two things can happen; either the neutron ends in the detector or it does not. This gives two pixels. At the next instrument setting, the same outcomes are possible. If the neutrons hit the detector, they are collected in pixel number 2, while neutrons missing are added to the missing neutrons from the previous setting. This goes on throughout the scan. Alternatively, if multiple detectors are used simultaneously, one splits all neutrons into the neutrons hitting individual detectors plus one for the neutrons that do not hit any detector. 

Although these two methods might appear to be completely equivalent, they are in fact not. Even though the end spectra seem equivalent there is one key difference. When all data points are measured at the same time, it is known that any neutron entering the instrument had the same probability distribution of being detected, and the total number of neutrons was fixed. When a single point at a time is being measured for a certain amount of time, or equivalently number of neutrons released from the source, it is not known that each detector setting had the exact same number of incoming neutrons, only that the total spectrum had a certain number. This is, albeit small, a difference between the two measurement styles. When the multi-detector instrument is used in a scanning setup the knowledge of the same total incoming neutron count is lost and one is to revert back to the same analysis as for the scanning setup. An example where these two setups are in use is a time-of-flight spectrometer measuring a powder sample and a single crystal. In the prior case noting is moved or scanned over during a spectrum acquisition while this is not the case for a single crystal. Here, usually the sample is rotated.

Looking at the case of many pixels being measured simultaneously, these are denoted $n_1,n_2,n_3,...,n_m$, such that there are $m$ different pixels. In addition, all neutrons not measured in these pixels (neutrons that do not reach any detector) are collected into $n_0$. That is, 
\begin{align}
\Delta N=\sum_{i=1}^m n_i, \\
N = \sum_{i=0}^m n_i.
\end{align}
Thus, in total $N$ neutrons hit the sample where $\Delta N$ of these hit the detectors and consequently $N-\Delta N$ hit outside of the detectors or are absorbed. The probabilities of a general neutron being detected in the individual pixels are denoted $p_i$, yielding
\begin{align}
\sum_{i=0}^mp_i=p_0+\underbrace{\sum_{i=1}^mp_i}_{\Delta p} = 1,\\
p_0 = 1-\Delta p.
\end{align}
It is these $p_i$'s that are of interest to the physical properties of the system and their values are correlated through the models of the scattering cross section. That is, in a simple case where the model is given by $p_i=A\mathrm{e}^{-\parh{\mu-x_i}^2/\parh{2\sigma^2}}+B$, i.e. a Gaussian peak on a flat background, the probabilities depend on each other through their $x_i$ position (which could represent $Q$ or $E$) and the model.

\paragraph{Multinomial Distribution}
In order to optimize these parameters, one needs to maximize the likelihood, which is given by a Multinomial distribution
\begin{align}
L = N!\prod_{i=0}^m\frac{p_i^{n_i}}{n_i!} = \frac{N!}{n_1!n_2!\cdots n_m!\parh{N-\Delta N}!}p_1^{n_1}p_2^{n_2}\cdots p_m^{n_m}\parh{1-\Delta p}^{N-\Delta N} \label{eq:Start}
\end{align}
By performing a Stirling's approximation and introducing the normalized quantities, see appendix \ref{app:Multinomial},
\begin{align}
q_i =& \frac{n_i}{\Delta N},\qquad \sum_{i=1}^m q_i = 1 \\
\tilde{p}_i =& \frac{p_i}{\Delta p}, \qquad \sum_{i=1}^m \tilde{p}_i = 1 \\
\end{align}
one can get to the log-likelihood
\begin{align}
    \ln{L}=& N\sbrac{\frac{\Delta N}{N}\sum_{i=1}^m\parh{q_i\ln{\frac{\tilde{p}_i}{q_i}}}+\frac{\Delta N}{N}\ln{\frac{\Delta p}{\frac{\Delta N}{N}}}+\parh{1-\frac{\Delta N}{N}}\ln{\frac{1-\Delta p}{1-\frac{\Delta N}{N}}}}.
\end{align}

The above log-likelihood is found when considering a collection of measurement data with a fixed total number of neutrons, i.e. $N$.

\paragraph{Poisson Distribution}
Taking one step back from the above derivation, what is usually performed is an analysis dealing with a data set where the total number of counts is not fixed, i.e. corresponding to the standard triple axis setup. This corresponds to removing the $p_0$ pixel. With this relaxation, the likelihood is given by the product of binomial terms for each detector, as

\begin{equation}
    P(D|M) = \prod_{i=1}^m P(D_i|M) = \prod_{i=1}^m p_i^{n_i}\parh{1-p_i}^{N-n}\frac{N!}{n_i!\parh{N-n_i}!},\label{eq:Binomial}
\end{equation}
where $n_i$ are the number of neutrons hitting the detector $i$, which has a probability of $p_i$, and the total number of neutrons are $N$. Taking this as a starting point, and going to the limit $n_i \ll N$, one radially finds the the likelihood to be a product of Poisson distributions\cite{Barlow}
 \begin{equation}
    P(D|M) \approx \prod_{i=1}^m\frac{\lambda^{n_i}\mathrm{e}^{-\lambda_i}}{n_i!},\label{eq:ProbabilityDataGivenModel}
\end{equation}
where $\lambda_i=Np_i$ is the average number of counts.
The largest possible probabilities are found when all $n_i \approx \lambda_i$. 
If the data is given as a vector of counts $n_i$ as a function of the index $i$, 
then using equation \eqref{eq:ProbabilityDataGivenModel} yields 
\begin{equation}
-\ln{P(D|M)}=\sum_{i=1}^m \sbrac{-n_i\ln{\lambda_i}+\lambda_i+\ln{n_i!}}\label{eq:ProbabilityOfData}.
\end{equation}
Now, applying a model to the data is equivalent to demanding that the ''true`` values, $\lambda_i$, follow a particular functional form
\begin{equation}
    \lambda_i = M_i(x_1, x_2, ...) = M_i(x_\alpha) ,
\end{equation}
where $x_j$ are the model parameters, shortened to the vector $x_\alpha$, that are to be optimized in the fitting procedure. Examples for data sets and fitting are given in section \ref{sect:simplemodels}. 

\subsection{Gaussian distribution}
It is instructive to compare by repeating the similar calculation for data governed by Gaussian statistics, which can be found in an expansion of the Poisson result (\ref{eq:ProbabilityDataGivenModel}) around a large value of $\lambda_i$\cite{Barlow}:
\begin{equation}
-\ln{P(D|M)}_{\rm Gauss} = \sum_{i=1}^m \frac{\parh{n_i-\lambda_i}^2}{2\sigma_i^2}-\frac{1}{2}\ln{2\pi\sigma^2}.
\label{eq:GaussLeastSquare}
\end{equation}
As the last term is independent of the model, $\lambda_i$, it merely represents a constant and is often removed. The same is true for the factor of 2 in the denominator of the first term. Maximizing the log-likelihood is thus equivalent to minimizing the quantity often denoted 
the {\em chi-square},
\begin{equation}
    \chi^2 = \sum_{i=1}^m \frac{\parh{n_i-\lambda_i}^2}{\sigma_i^2} .
    \label{eq:ChiSq}
\end{equation}
The whole procedure of minimizing this equation is often known as {\em least squares fitting} \cite{Barlow}.
However, applying this Gaussian statistical treatment, a relation between $\sigma_i$ and the intensity is needed. 


\subsection{Fitting experimental data}
Fitting a model using the above found likelihoods then consists of optimizing $\ln{L}$ where the model parameters, $x_\alpha$, give the values for $p_i$ or $\lambda_i$. It is important to note that only the dependence of the log-likelihood on these parameters matters; everything else is constant and can be discarded. 

The Multinomial log-likelihood can be split into two parts; one concerning the zeroth pixel, the other the rest. The parameters only changes the latter part, which can be found to be proportional to
\begin{equation}
-\ln{P(D|M)} = -N\sum_{i=1}^m q_i\ln{\tilde{p}_i},\label{eq:MultinomialLL}
\end{equation}
see appendix \ref{app:Multinomial} for details.

For the Poisson distributed data, the negative log-likelihood contains the term $\ln{I_i!}$ which is independent of the model parameters. In effect, one has to optimize
\begin{align}
    -\ln{P(D|M)} = \sum_{i=1}^N\sbrac{-n_i\ln{\lambda_i}+\lambda_i}.
\end{align}
Comparing the two above log-likelihoods, they are almost equivalent except for the $\lambda_i$ term and only normalized terms in eq.~\eqref{eq:MultinomialLL}. This is exactly the difference between the two measurement techniques; for the term in the Poisson $\sum_{i=1}^n \lambda_i = \sum_{i=1}^n N p_i = \Delta N$, but there is no constraint on $\Delta N$ relative to $N$. In the Multinomial case, a term $\sum_{i=1}^n \tilde{p}_i$ was present, but is known to always sum to unity.

Lastly, for the Gaussian distribution the log-likelihood is simply proportional to $\chi^2$ and does thus not need to reformulated.

In the Gaussian log-likelihood, it would also be possible to use the model value for the uncertainty, {\em i.e.} $\sigma_i = \sqrt{\lambda_i}$. However, a lot of computational flexibility ({\em e.g.}\ in normalization and background subtraction) is gained if $\sigma_i$ can be determined in a model-free way, {\em i.e.}\ directly from the individual data point. Hence, the approximation used in almost any fitting program is $\sigma_i \approx \sqrt{n_i}$.

Difficulties arise from using this equation in the extreme low-count limit. In particular, when a counting number of 0 is measured, we have $\sigma_i = 0$, corresponding to a (physically unreasonable) zero uncertainty on the data point. Statistically, this would mean that it is known with certainty that the true value, $\lambda_i$, equals zero. 
This will, in turn, result in the model fits being forced through zero at these points.

For these reasons, practical applications of modeling of scattering data use different tactics to accommodate zero count values. 
The most often used way to circumvent the zero-count problem is by increasing the uncertainty of the zero-measurement to unity\cite{Mantid,iFit}. This, however, allows for the unphysical situation where $\lambda_i$ is just as likely to be positive as negative. One could device another method where zero-measurements are removed altogether. This of course introduces a strong bias, as measuring a point with zero counts contains a lot of information being ignored. Alternatively, one can shift the intensity of zero counts to 0.5 and use this value also as the uncertainty. Table~\ref{tab:ZeroCountTactic} show these three different tactics. 

\begin{table}[h!]
    \centering
    \begin{tabular}{c|c|c}
        Method & Intensity$_0$ & $\sigma_0$ \\ \hline
        BG1 & 0 & $1$ \\
        BG2 & 0.5 & $0.5$ \\
        BG3 & Remove & Remove
    \end{tabular}
    \caption{Three different tactics for dealing with zero count values in Guassian statistics. The third method simply removes zero count points.}
    \label{tab:ZeroCountTactic}
\end{table}

In the similar case, when a count of unity is found, the corresponding uncertainty is then $\sigma_i = 1$. This means that a negative value is only '1 $\sigma$ away' corresponding to the true value being positive with a probability of 84.1\%, leaving an almost 16\% probability of it being negative - which is again unreasonable. We do not here consider modifications of the errorbar of count values of 1. However, we can state that if a data set contains many low-count numbers, the use of Gaussian statistics is certainly imprecise. 

In the rest of this paper, we will quantify how these introduced imprecisions affect the data analysis in a few simple examples, where we also compare with the more accurate Poisson and multinomial treatments.

\section{Fits of two simple Models} \label{sect:simplemodels}

We here set out to investigate the difference between minimizing the three different log-likelihoods, when used on simple, synthetic counting data. We first show a study of a data set of no features, i.e. a flat background. Later, we discuss the case of one simple Gaussian peak on a flat background. 

For the flat background, 1000 individual spectra are generated using the \texttt{numpy.random.poisson} method, implemented in Python \cite{Numpy}, where the mean count is calculated from the model. All $x_i$ lie within -1 to 1. For the peak shape on a flat background 10\,000 individual spectra with a total of 1000 counts in each, with once again $x_i$-values between -1 and 1. As the total neutron count is fixed, the spectra are generated by the \texttt{numpy.random.multinomial} method.
In both cases, each spectrum is fitted using a) the Multinomial log-likelihood, b) the Poisson log-likelihood and, c) the Gaussian least squares method, where $\sigma_i = \sqrt{I_i}$ for a series of model parameters. In the latter case, the three different tactics of dealing with zero counts from Table~\ref{tab:ZeroCountTactic} were used in turn. In order to ensure physical convergence, a bound has everywhere been imposed on the background and amplitude variables, $B$ and $A$, so that $B \geq 0$ and $A \geq 0$.

\subsection{Constant Background}\label{sec:ConstantBackground}
We here consider the simplest model
\begin{equation}
    \lambda_i(x_i|B) = B.
\end{equation} 
We estimate the background value, $B_{\rm est}$ for true $B$ values lying in the range 0 to 20, using the different schemes discussed above and 21 data points per series. As the Multinomial log-likelihood requires an optimization of the normalized probabilities, there remain no parameter to fit. Thus, the Multinomial log-likelihood is not fitted to the featureless background. 

\begin{figure}[h!]
    \centering
    \includegraphics[width=0.45\linewidth]{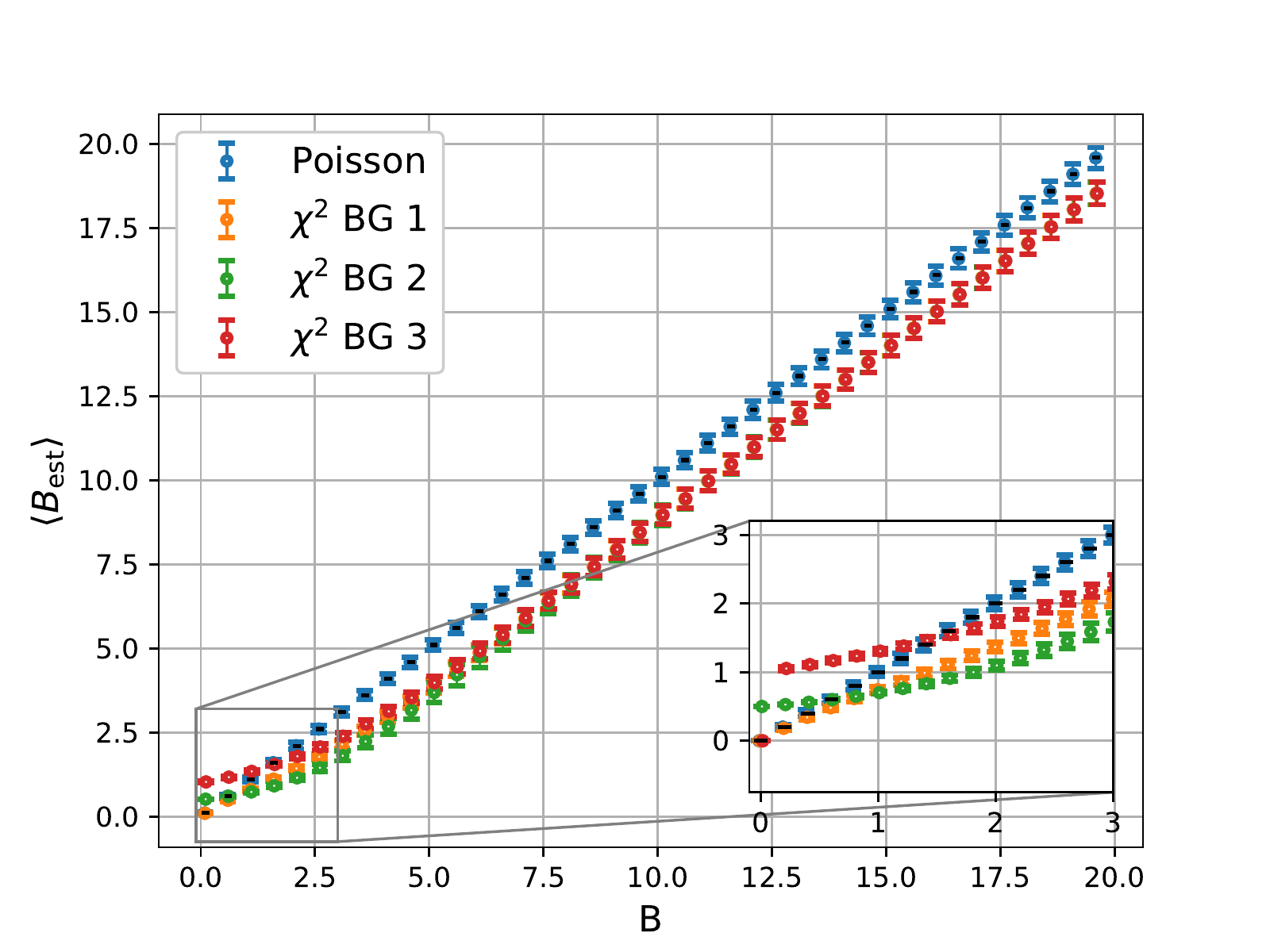}
    \includegraphics[width=0.45\linewidth]{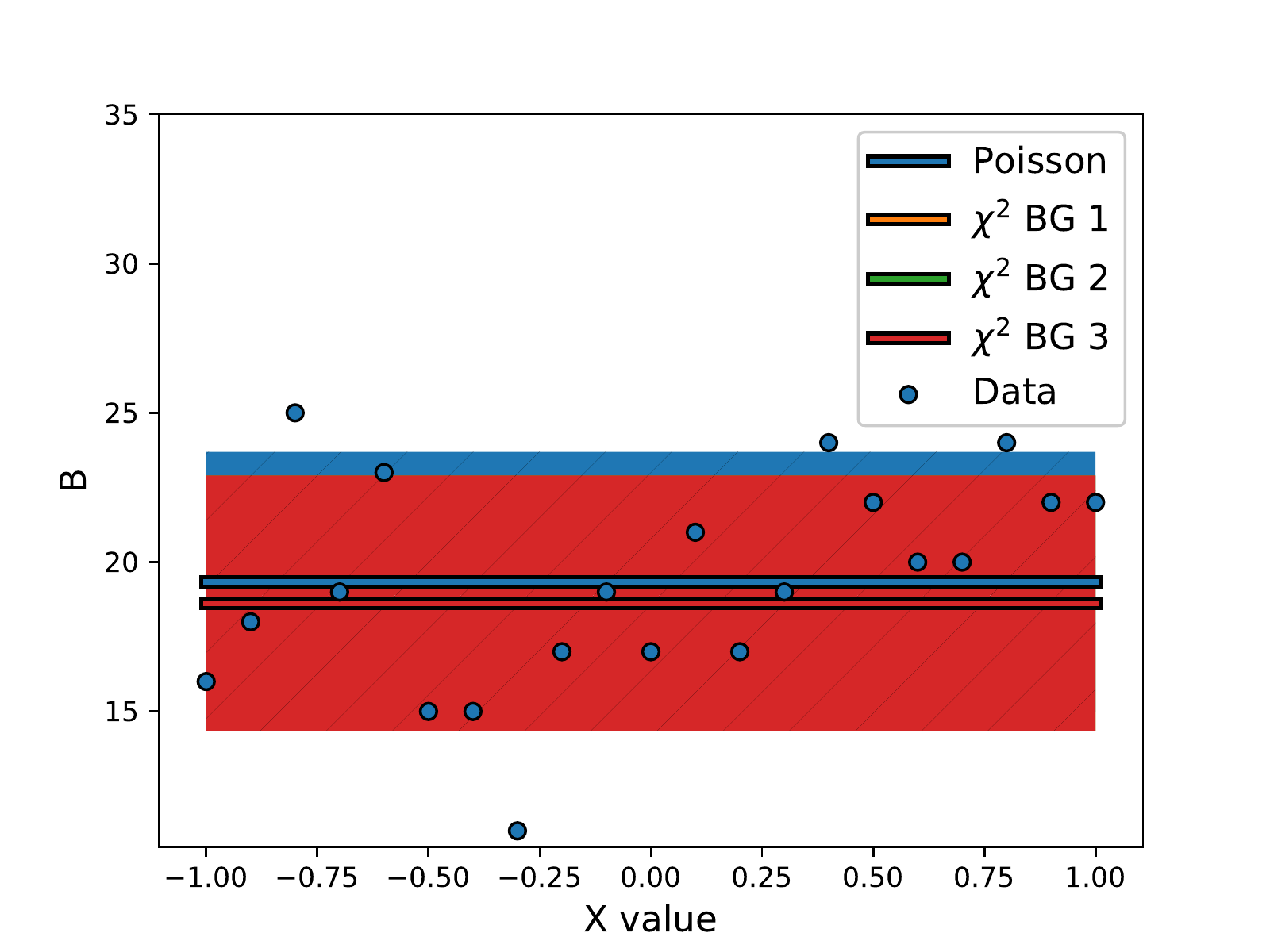}
    \caption{\textbf{Left}: Average estimated background value obtained for fits to 1000 random data sets, fitted with the Poisson log-likelihood method and the three least-square methods, as explained in the text. Black lines signify the true parameter value, $\lambda_i$.  The standard deviation on the background estimations are shown as an ''errorbar`` on each point.
Note that the standard deviation of the plotted mean value is therefore a factor $\sqrt{1000}$ lower than the plotted standard deviation. \textbf{Right}: Example of a single dataset for background value of $B = 20$ with confidence interval of $1 \sigma$ is shown for all fits, as explained in sec.~\ref{sec:Uncertainty}. All least square methods result in the same fit and uncertainty area.}
\label{fig:FlatBackground}
\end{figure}

Fig.~\ref{fig:FlatBackground} shows the mean estimation parameter and the standard deviation on it, the size of which is described in sec.~\ref{sec:Uncertainty}. 
While the Poisson fit shows a striking agreement with the underlying model, we observe a clear underestimation of the background parameter in all the Gaussian least square method fits. This is visible for all medium and large values of the background parameter, $B \geq 1.5$ , {\em i.e.} also outside the Poisson regime. This is a feature of Gaussian statistics, caused by the fact that lower count numbers are ascribed smaller error bars and therefore have higher relative weights in the chi-square fit \eqref{eq:ChiSq}. At large values of $B$, it can be shown that the deviation tends to a constant $\Delta B = \langle B_{\rm est} \rangle - B = -1$, see appendix \ref{app:BackgroundEstimators}. This means that the relative error, $\Delta B/B \sim 1/B$ and thus still around 10\% when we leave the Poisson regime.

Turning to the case of low background rates, $B \leq 1$, we note that the Gaussian methods produce larger results than the true value ($\Delta B > 0$) for methods BG2 and BG3; with the worst results coming from BG3. This overestimation of the background as compared to the other chi-square fits is a natural consequence of modifying the zero-count observations.   

In the medium range $B = 1.5$ to 3.0, the methods BG1 and BG2 are systematically too low, and here BG3 becomes more precise. We note that the BG2 method (setting zero counts to 0.5 and the same value for the corresponding error) is everywhere worse than the BG1 method (setting the zero-count error bar to unity). The BG3 method (ignore zero counts) is found to be the most precise method in the range $1 \leq B \leq 6$. Overall, however, for the flat background case the BG1 method can be judged to be the best of the Gaussian methods across all scales of background amplitude. This result justifies the frequent use of the BG1 tactics.

None of the Gaussian methods, however, compare anywhere near the Poisson method in fitting precision for this simplest of models.

\subsection{A single Gaussian peak on a constant background}
An example very relevant to scattering is that of a peak with the shape of a Gaussian on a constant background. This model is given by
\begin{equation}
    \lambda_i (x_i|A,\mu,\sigma,B) = A\mathrm{e}^{-\frac{\parh{x_i-\mu}^2}{2\sigma^2}} +B, \label{eq:GaussianPeak}
\end{equation}
where $A$ is the amplitude of the peak, $\mu$ is the mean value, $\sigma$ is the peak width (which should not be confused with the statistical standard deviation of the counting data), and $B$ is the constant background. 

In contrast to the above data, it here makes sense to also do parameter optimization using the Multinomial log-likelihood, where
\begin{align}
\tilde{p}_i = \frac{p_i}{\sum_{j=1}^m p_j} =& \frac{A\mathrm{e}^{-\frac{\parh{x_i-\mu}^2}{2\sigma^2}}+B}{\sum_{j=1}^m A\mathrm{e}^{-\frac{\parh{x_j-\mu}^2}{2\sigma^2}}+B} \\
=& \frac{\frac{A}{B}\mathrm{e}^{-\frac{\parh{x_i-\mu}^2}{2\sigma^2}}+1}{\sum_{j=1}^m \frac{A}{B}\mathrm{e}^{-\frac{\parh{x_j-\mu}^2}{2\sigma^2}}+1},
\end{align}
which is a model only depending on three parameters: $A/B$, $\mu$, and $\sigma$, and not the four present in the other fitting schemes.

\begin{figure}[ht!]
    \centering
    \includegraphics[width=0.49\linewidth]{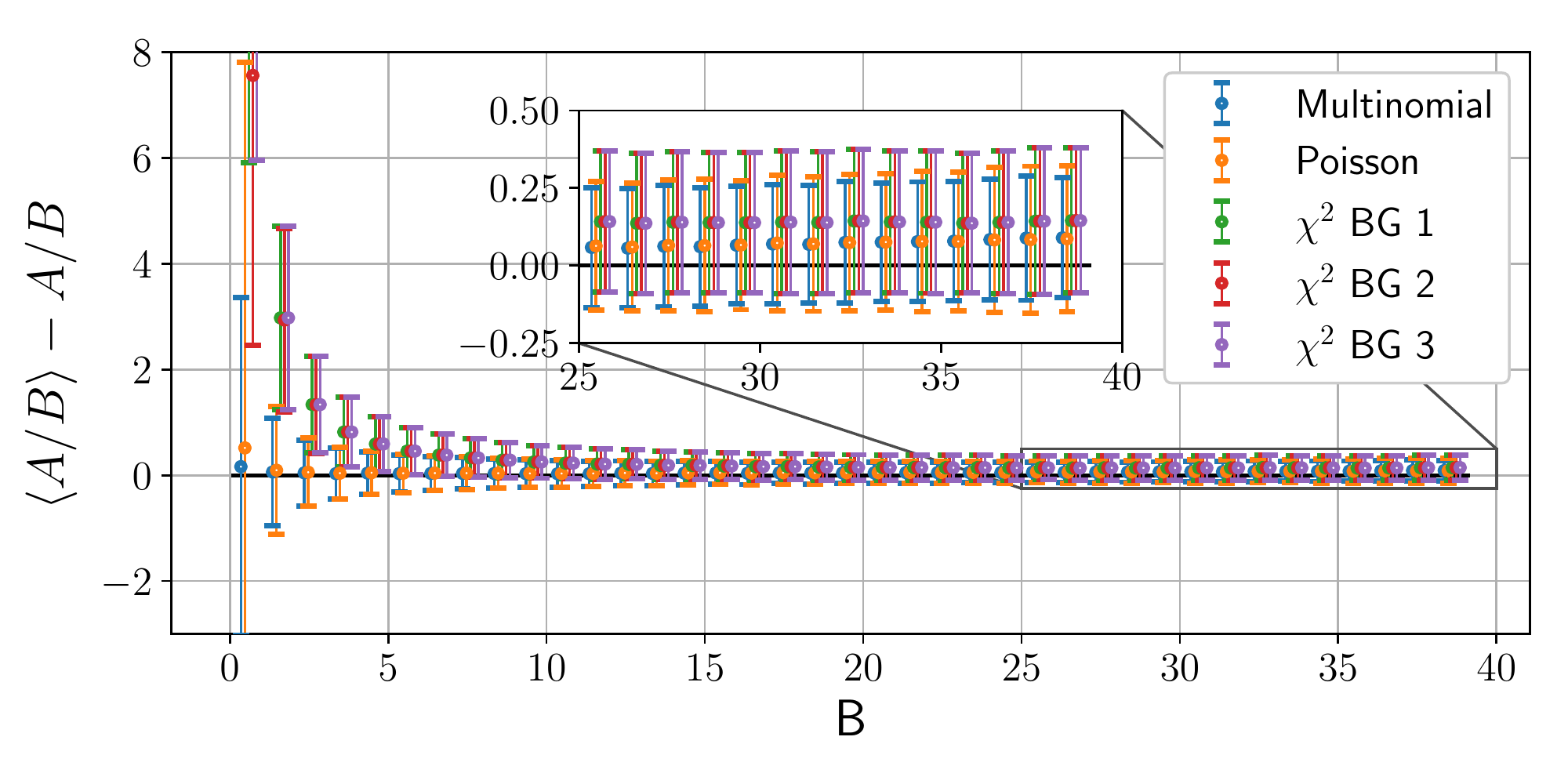} 
    \includegraphics[width=0.49\linewidth]{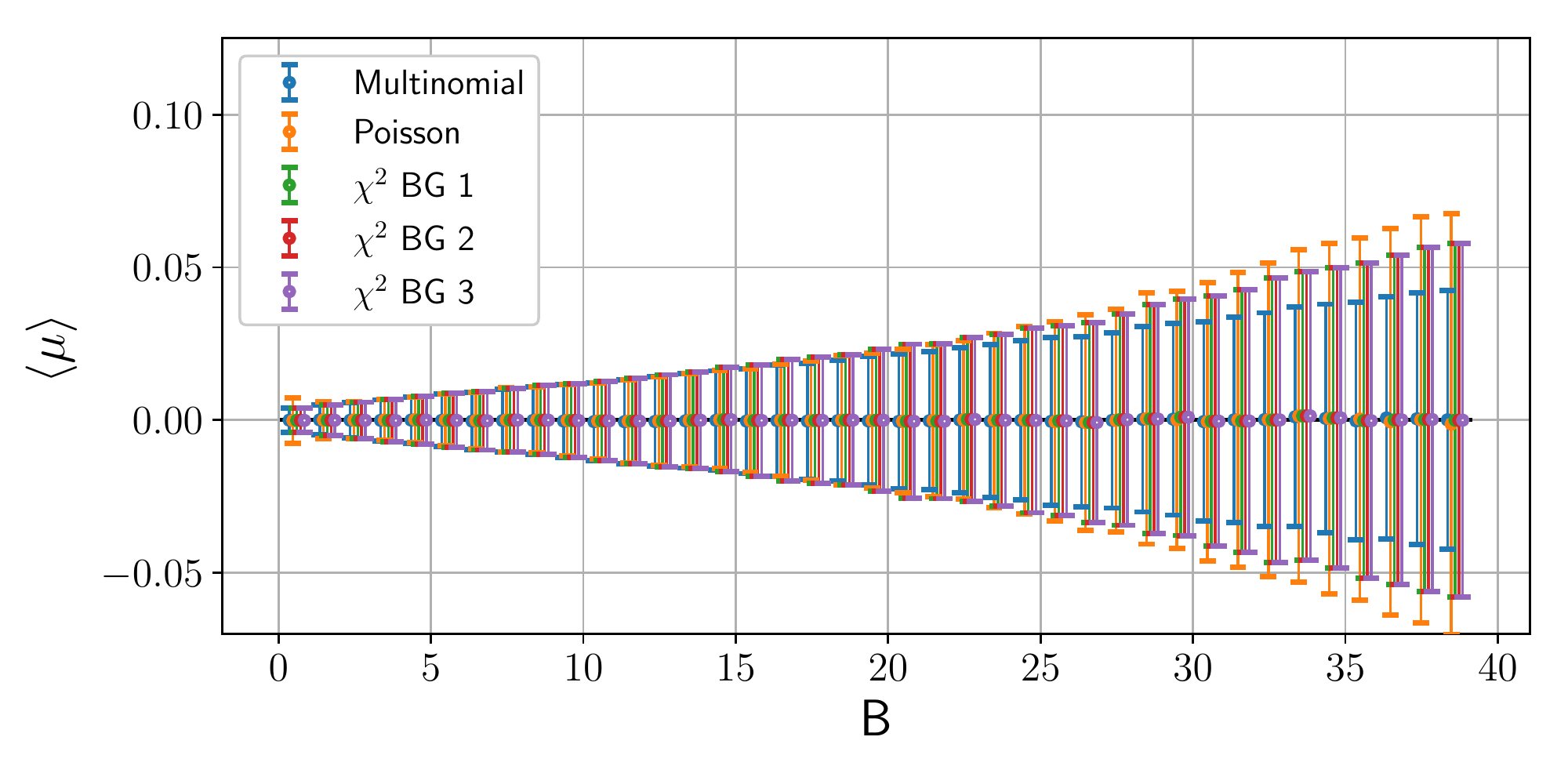} 
    \includegraphics[width=0.49\linewidth]{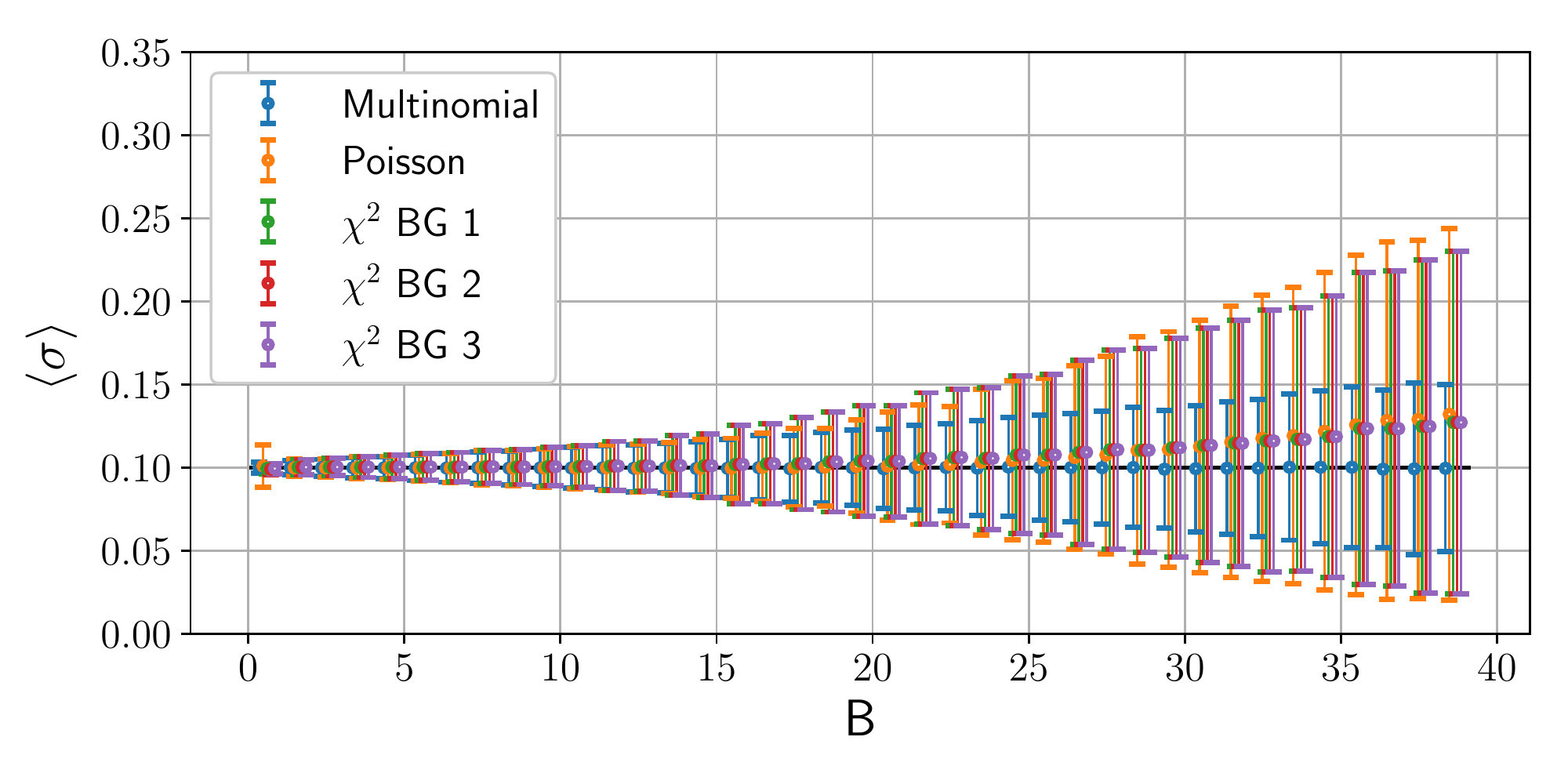} 
    \includegraphics[width=0.49\linewidth]{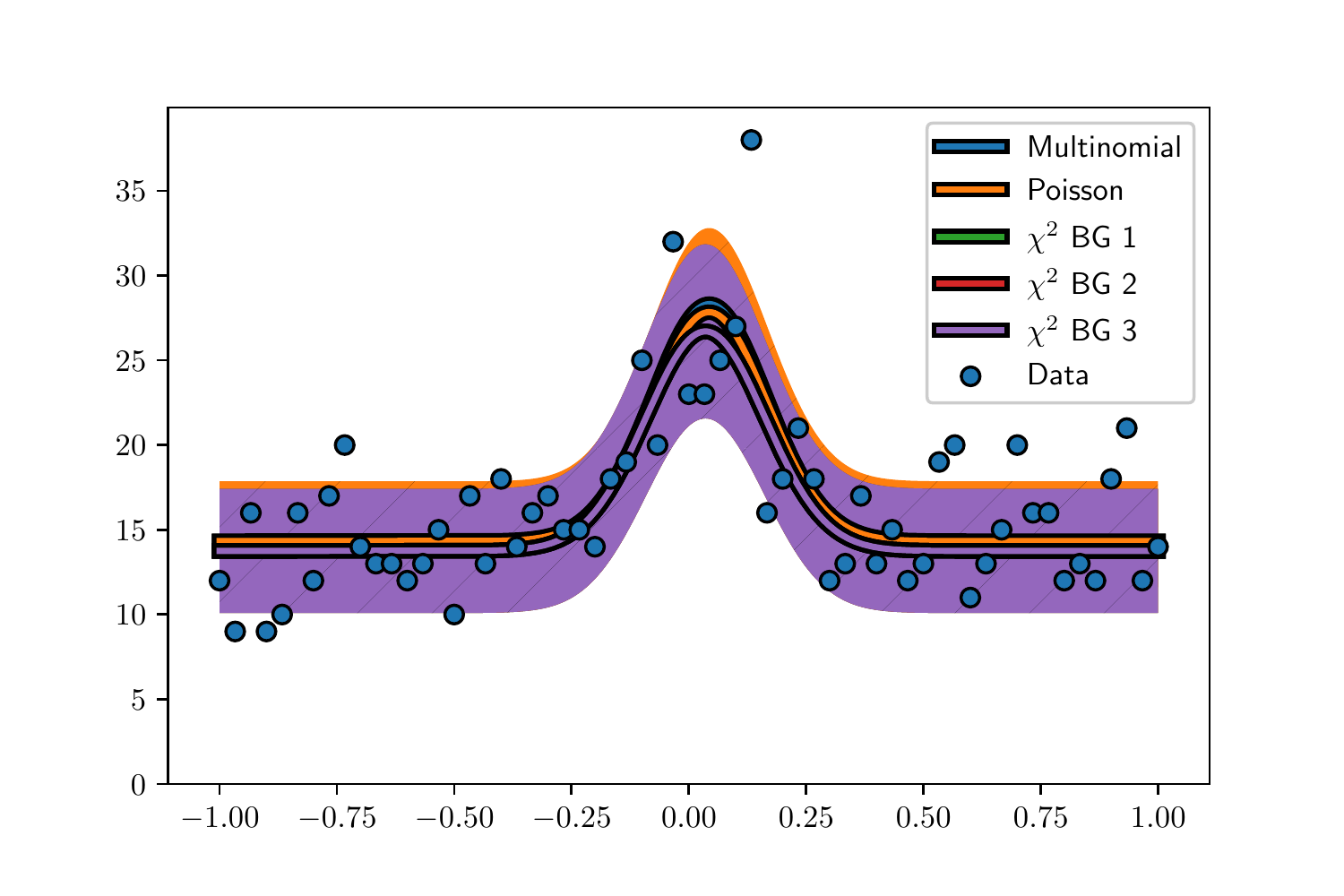} 
    \caption{Distance away from the truth for estimated values of $A$/$B$, $\mu$, $\sigma$, 
    and a sample data set with a background level $B=21.1$. This background gives a signal to noise of about 1 which with 1000 counts over 61 pixels result in about 14 counts in background and 28 in signal. 
    }
    \label{fig:PeakCollection}
\end{figure}

For this model the different values of the background, $B$, was used with fixed center, $\mu=0$, amplitude, $A=20$, width, $\sigma=0.1$, and range of $x$ values, $N=61$ points from $x=-1$ to $x=1$. That is, although the background level is set to e.g. 15, the requirement of a maximum of 1000 counts in spectrum still apply. That is, background and amplitude levels do not correspond directly to mean count numbers but rather to relative intensities. One typical data set is shown in Fig.~\ref{fig:PeakCollection}, as well as the statistics of the fits of the five models to each of the synthetic spectra. Their corresponding confidence intervals, corresponding to 68.27 \% are plotted on top. As the Multinomial log-likelihood does not provide an amplitude measure directly, the distance away from the true $A/B$ is plotted, see top left of Fig.~\ref{fig:PeakCollection}.

In the data, we observe the same tendency of the three Gaussian least square fits overestimating the $A/B$ parameter. Especially at small values of the background they diverge substantially from the true value. For larger values, an offset seems to be present, which could be expected from the analysis of the optimal fitting parameters for the feature-less fit in sec.~\ref{sec:ConstantBackground}. Both the Poisson and Multinomial methods are quite accurate at small values of background and continue to be up to a background value of around 25. Above this, there seem to be a constant offset for all higher background values. This is, however an artefact of the limited total count in the spectrum. If this is increased to 10\,000 only the Poisson does not improve its mean, while the Multinomial does. Increasing the total counts also reduces the errorbars as expected, not shown.


When it comes to the center position of the peak, $\mu$, all methods agree across all levels of background, accurately finding the true mean value, $\langle \mu_{\rm est} \rangle = 0$, as could be expected from symmetry arguments - the counting error is treated equally for positive and negative values of $x_i$. All methods share a general trend of larger standard deviation of the estimator for larger values of background, simply reflecting the larger level of noise on each data point when a background is added. Especially the Multinomial method has small uncertainties on the mean as compared to the other methods. Around the background value of $\sim$ 22 the size of the errorbars from the Poisson method start to increase in size. In contrast, the error of the Multinomial method only grows slowly for increasing background values. 

There is no doubt that the Multinomial method out-performs the other 4 methods when the peak width is to be determined. When the background to amplitude reaches a ratio of 1, all but the Multinomial method, on average, overestimate the peak width. The Multinomial method not only finds the correct width but also with significantly smaller standard deviation.

\section{Model normalization, visualization, and uncertainty}\label{sec:Uncertainty}
In most neutron experiments, the acquired raw count is somehow normalized. Often this is done with respect to monitor count, resolution volume and detector sensitivity, just to mention a few. The standard progression is to normalize the intensity measured and then find the estimated uncertainty on the data points, that is
\begin{equation}
    I_i \rightarrow \frac{I_i}{N_i}, \qquad \sigma_i=\sqrt{I_i} \rightarrow \frac{\sqrt{I_i}}{N_i}.
\end{equation}
This introduces a further uncertainty on the counting number from the measurement of the monitor value. By applying the error propagation by adding their uncertainties in quadrature one gets
\begin{equation}
    \sigma_i = \sqrt{\frac{I_i}{N_i^2}+\frac{I_i}{N_i^3}}.
\end{equation}
As the monitor count is often orders of magnitude above the actual detector counts, we here drop the latter term. 

However, the cleanest way to perform this transformation is to transform the model instead of the data, for example that the expected count rate is proportional to the count time:
\begin{equation}
\lambda_i \rightarrow \lambda_i \alpha_i ,
\end{equation}
where $\alpha_i$ is the (point dependent) normalization constant.
In a least-square fit, the value of the variance weighted square deviations, $\chi^2$, will now be given as
\begin{equation}
    \chi^2 = \sum_i \frac{\parh{\lambda_i-I_i}^2}{\sigma^2_i}\qquad \rightarrow \qquad \sum_i \frac{\parh{\lambda_i\alpha-I_i}^2}{\sigma^2_i} =  \sum_i \frac{\parh{\lambda_i-\frac{I_i}{\alpha}}^2}{\frac{\sigma^2_i}{\alpha^2}}.
\end{equation}
Thus, transforming the model by $\alpha_i$ is identical to the transformations on the individual data points: $I_i\rightarrow I_i/\alpha_i$ and $\sigma_i\rightarrow\sigma_i/\alpha_i$. This is the way that this normalization is usually implemented in practice, when using Gaussian statistics. However, when the Poisson log-likelihood method is used, the normalization belongs only to the model, since a scaling of the number of counts will interfere with the Poisson counting statistics. 
The Multinomial method, on the other hand, re-normalizes all of the data, in contrast to the Poisson, such that the absolute scale is irrelevant. But this is only the case for an overall scaling as relative normalizations between data points still has to be taken into account, as is the case for the $\chi^2$ methods. This lack of absolute scale also impacts the visualization of the result. Plotting the optimal parameters on top of the fitted data requires the scale of the data to be found. It is, however, simply given as the sum of all counts and a re-scaling is trivial.

When visualizing data, it is common practice to display an errorbar on the individual data points, representing the statistical uncertainty. However, following the discussion on the Gaussian and Poisson statistics above, this is formally a wrong presentation of counting data. In principle, there is no uncertainty on the actual measurement in a given point. Rather, the uncertainty lies on the estimation on the underlying true scattering intensity $\lambda_i$, in other words: on the model parameters. 
With this in mind, a more statistically consistent way of visualizing data would be to show data points without errorbars, while showing the refined models with ''error intervals``, which could be shaded areas corresponding to the regular 1$\sigma$ confidence interval. This is the method we have used to display our data above, Fig~\ref{fig:PeakCollection}. The way of visualizing error does of course not change the underlying analysis, {\em e.g.} the estimation of model parameters, but is merely a visual change. Further, it also highlights the fact that the model extrapolates from the data fitted and predicts the true hind-lying $\lambda_i$ for all possible values of $x_i$ despite only a limited number of $x_i$ values has been observed. 

Finding the confidence intervals for the Multinomial and Poisson methods requires a little work. 
By the notion of uncertainty on the model it is meant that it is independent of the uncertainty on the fitted parameters and represents the statistical uncertainty in drawing counts from its distribution. This is an alternative to providing an error estimate on the data points with the best fitting model plotted on top.  
In the case of the least squares fit, the region of model uncertainty is the count values corresponding to $\pm 1\sigma$ as found from solving
\begin{equation}
    \int_{-a}^0 \frac{1}{\sqrt{2\pi}}\mathrm{e}^{\frac{-x^2}{2}}=\alpha \qquad \mathrm{and} \qquad \int_0^b \frac{1}{\sqrt{2\pi}}\mathrm{e}^{\frac{-x^2}{2}} = \beta,
\end{equation}
for $a$ and $b$ with $\alpha=\beta=0.3173$, yielding the usual $a=b=1$. However, for the Poisson log-likelihood statistics, the corresponding procedure is less obvious, in particular due to the discrete nature of the Poisson statistics. A number of different approaches have been discussed in literature \cite{Patil2012} that both cover the wanted area as tightly as possible and without skew. The main discussion issues are 1) Does the error estimate have to be integer or can it be relaxed to be non-integer? 2) Should the range covered above and below the mean value be symmetric?  In the present case of scattering data, we will usually normalize the underlying model with (at least) the counting time or the monitor counts, thus allowing for a loosening of the discrete nature. Regarding skewness, it is of greater scientific value to have a statistically true representation than an aesthetically pretty figure. 

Thus, one can define the confidence interval limits $a$ and $b$ equivalently as for the Gaussian with 
\begin{equation}
    \int_a^\lambda \frac{e^{-\lambda}\lambda^x}{x!} d\,x = \alpha \qquad \mathrm{and} \qquad \int_{\lambda}^b \frac{e^{-\lambda}\lambda^x}{x!} d\,x = \beta ,
\end{equation}
for $\alpha=\beta=0.3173$ corresponding to the integral of a Gaussian from $0$ to $\sigma$. An example is shown in Fig.~\ref{fig:PeakCollection}. As the Poisson distribution is skewed so are the values of $a$ as compared to $b$.

For the Multinomial distribution, one can use the confidence interval methods for binomial distribution. At each point along the fitted curve the success probability is simply the estimated value, while all other outcomes are regarded as fails. As was the case for the Poisson, many different procedures for calculating confidence intervals exist \cite{Wallis2013}. Weighing calculational complexity and correctness, it has been chosen to use the Wilson score interval with continuity correction. Specifically, the confidence interval is found from 
\begin{align}
CI_{\rm upper}(p,n,z) &= \min\left\{\frac{2np+z^2+z\sqrt{z^2-\frac{1}{n}+4np(1-p)-(4p-2)}+1}{2(n+z^2)},1\right\}\\
CI_{\rm lower}(p,n,z) &= \max\left\{\frac{2np+z^2-z\sqrt{z^2-\frac{1}{n}+4np(1-p)+(4p-2)}+1}{2(n+z^2)},0\right\},
\end{align}
for a total of $n$ counts and $z$ is the probit corresponding to the wanted confidence interval. In the case of 1$\sigma$, $z = 0.952$.

\section{Error estimate on parameters}
An experimentally determined parameter has little scientific value without a corresponding uncertainty value. That is to say that when tabulating fitting parameters or other extracted variables, one needs to quantify the degree to which this value represents the true underlying numerical value. 
In general, two different ways of estimating the error exists; 1) change only the parameter in question until the log-likelihood value changes a certain amount or 2) change the parameter in question and optimize the others until the log-likelihood has changed by the given amount. 

For the case of a normally distributed variable being fitted by a single parameter, the uncertainty on the parameter is given by a change in the chi-square value of unity, or, when the log-likelihood method is used, by a change of this value by 0.5. This, in turn, corresponds to a confidence interval of 68.27\%, usually denoted the $1\sigma$ interval \cite{Barlow,James}. However, in the multi-dimensional case with many parameters,
a change of 0.5 in the log-likelihood no longer represents the $1\sigma$ interval. Instead, the task is to find the inverse of the cumulative density, such that the 68.27\% confidence interval is found. All of this has already been implemented in the software package \texttt{Minuit} \cite{Minuit}. The two above described methods of acquiring the uncertainties still apply and both of these are available in \texttt{Minuit}; one through the regular minimization and one by the \texttt{minos} algorithm. This method is computationally heavier and will in general return non-symmetric errors. 

\begin{figure}[!ht]
    \centering
    \includegraphics[width=0.75\linewidth]{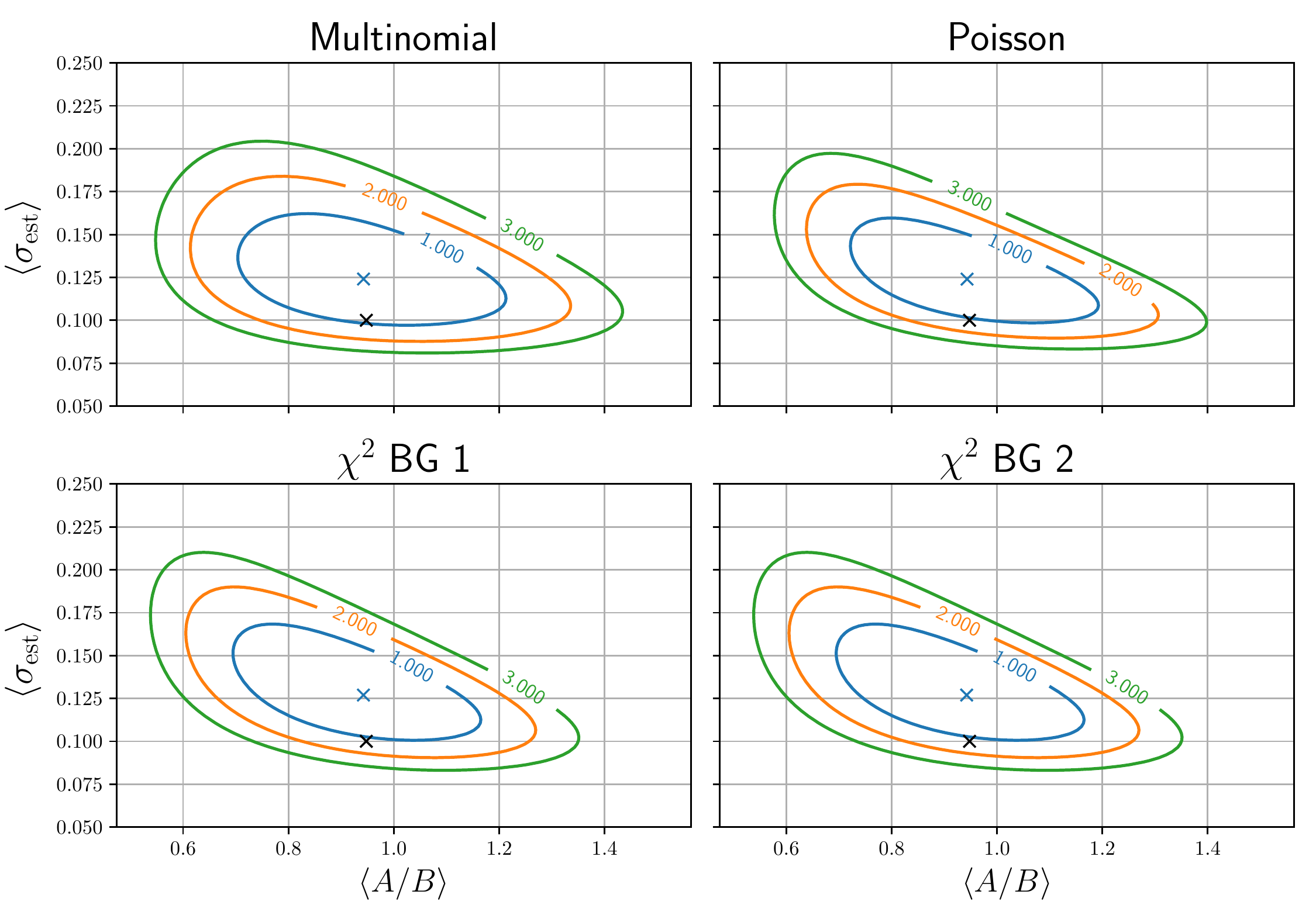}
    \caption{Estimated $1\sigma$, $2 \sigma$, and $3 \sigma$ confidence intervals for the signal-to-background value ($A/B$) and peak width ($\sigma$) fitting parameters for the synthetic data set in Fig.~\ref{fig:PeakCollection} of a Gaussian peak on constant background as described in section~\ref{sect:simplemodels}. The value of the true model parameters are displayed as a black dot in each panel at the point $(A/B,\sigma) = (0.926,0.1)$. 
    }
    \label{fig:ErrorEstimateMinuit}
\end{figure}

Further, curves for constant log likelihood can also be plotted, and an example for the Multinomial, Poisson and $\chi^2$ for error scheme 1 and 2 are shown in Fig.~\ref{fig:ErrorEstimateMinuit}, for the template data in Fig.~\ref{fig:PeakCollection}. That is, the signal-to-background level is $20.0/21.6 \approx 0.93$ and $\sigma$ is 0.1. Error scheme 3 is not shown as it completely resembles scheme 1 and 2. The constant log likelihood curves are plotted as function of estimated $A/B$ and $\sigma$. Naïvely one would conclude that the Multinomial confidence interval is larger than those for the other methods, but what is not taken into account is the uncertainty for these in the determination of $B$. For this effect to be visible, multiple different spectra are to be generated and fitted.

Looking at the distribution of parameter estimations for the three different statistics types it is seen that on average the Multinomial distribution is both most accurate and precise with the Poisson statistics following its precision. Comparing the extend of the 1, 2, and 3 $\sigma$ intervals for the Multinomial distribution with the error estimate in Fig.~\ref{fig:ErrorEstimateMinuit} it can be argued that its error estimate is too large. A true correspondence between change in log-likelihood and error estimate might not have been achieved resulting in an overestimation of uncertainty in single parameters.

\begin{figure}[!ht]
    \centering
    \includegraphics[width=0.45\linewidth]{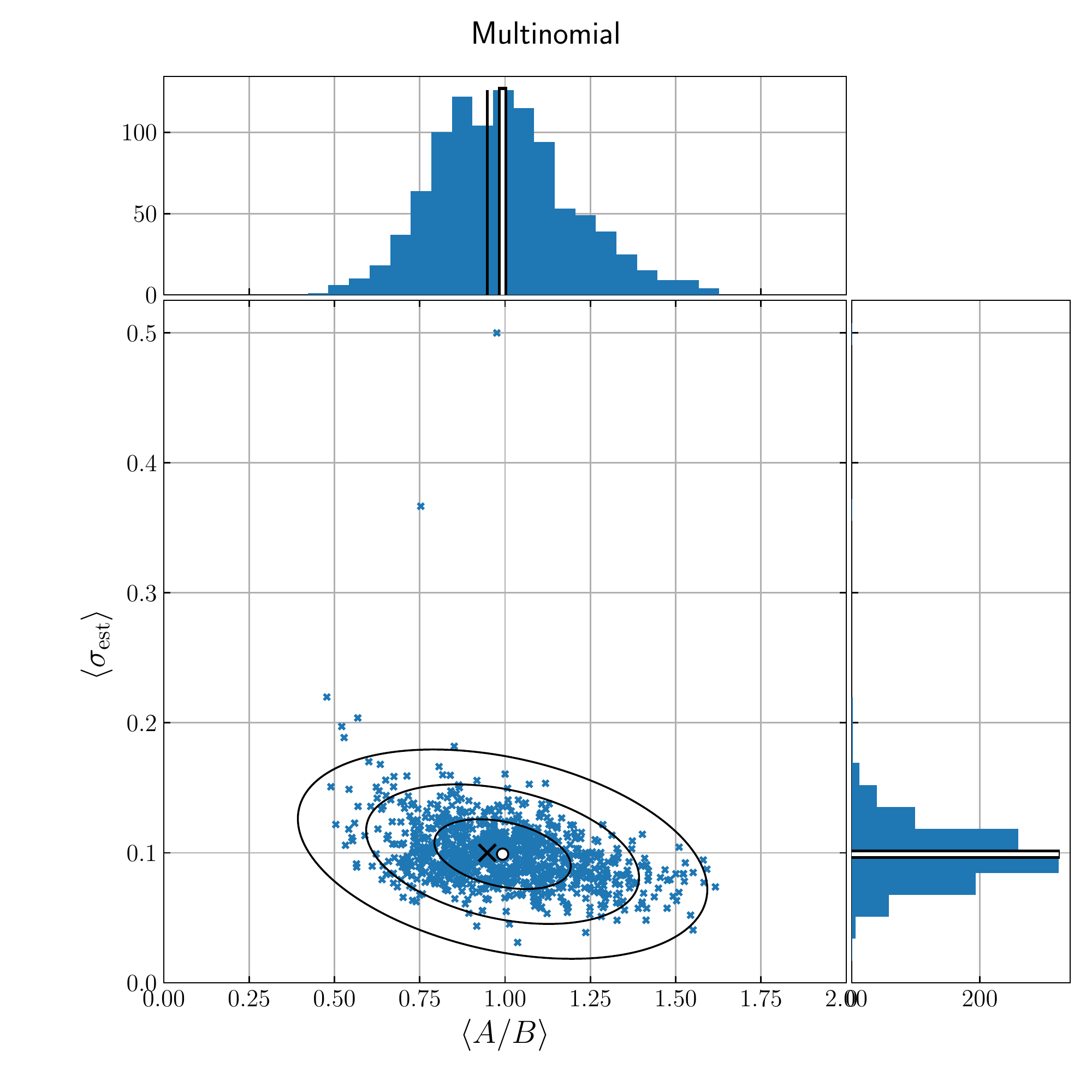}
    \includegraphics[width=0.45\linewidth]{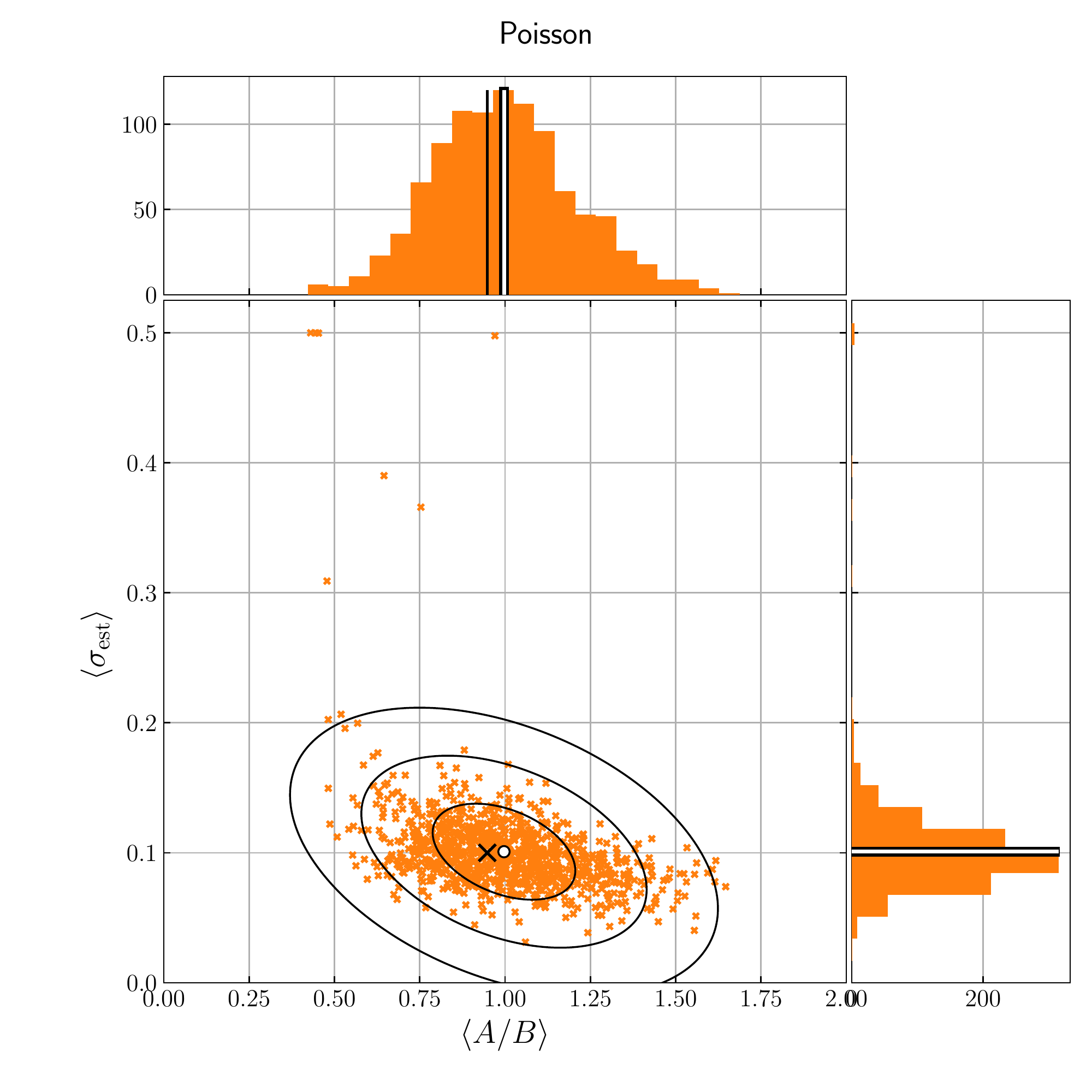}
    \includegraphics[width=0.45\linewidth]{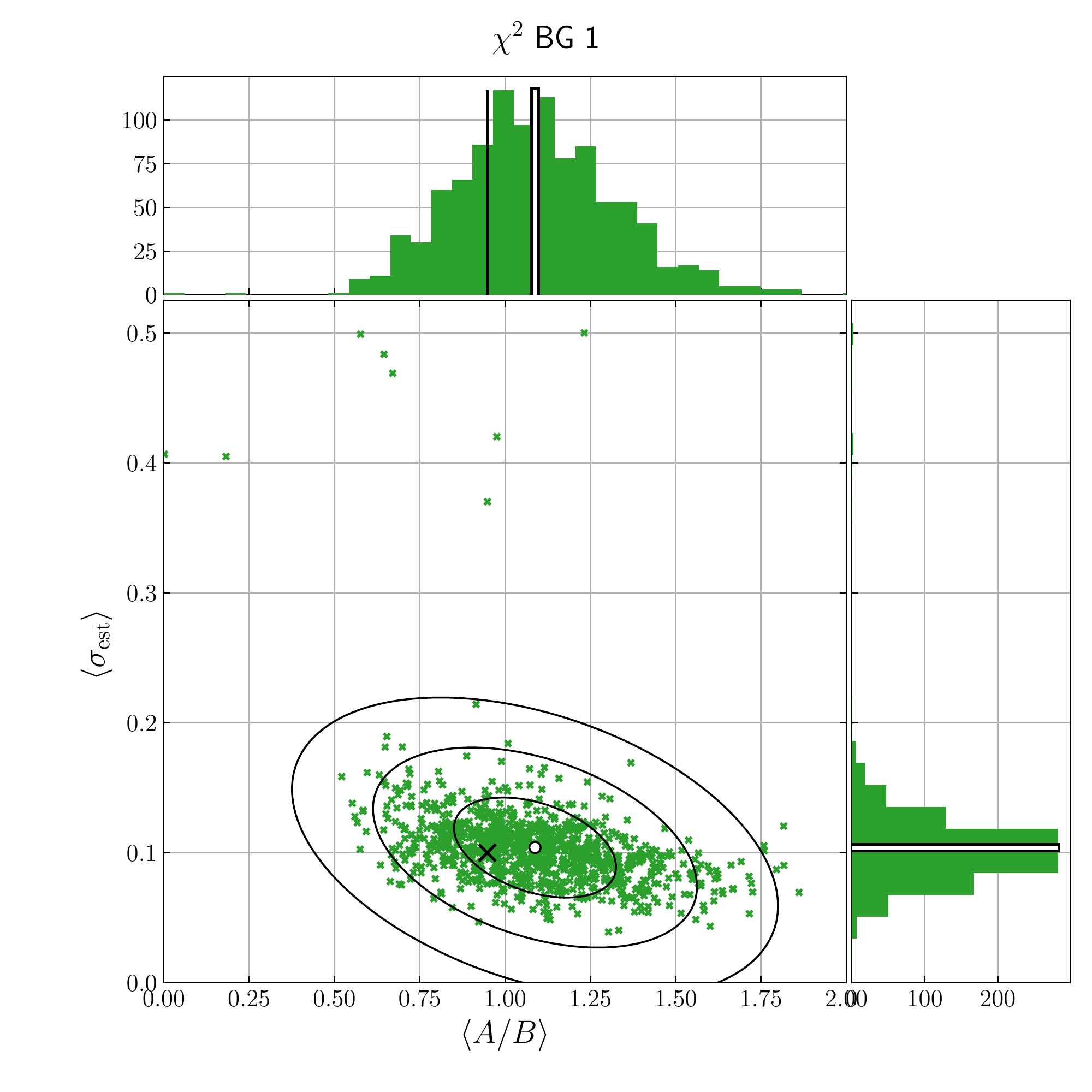}
    \includegraphics[width=0.45\linewidth]{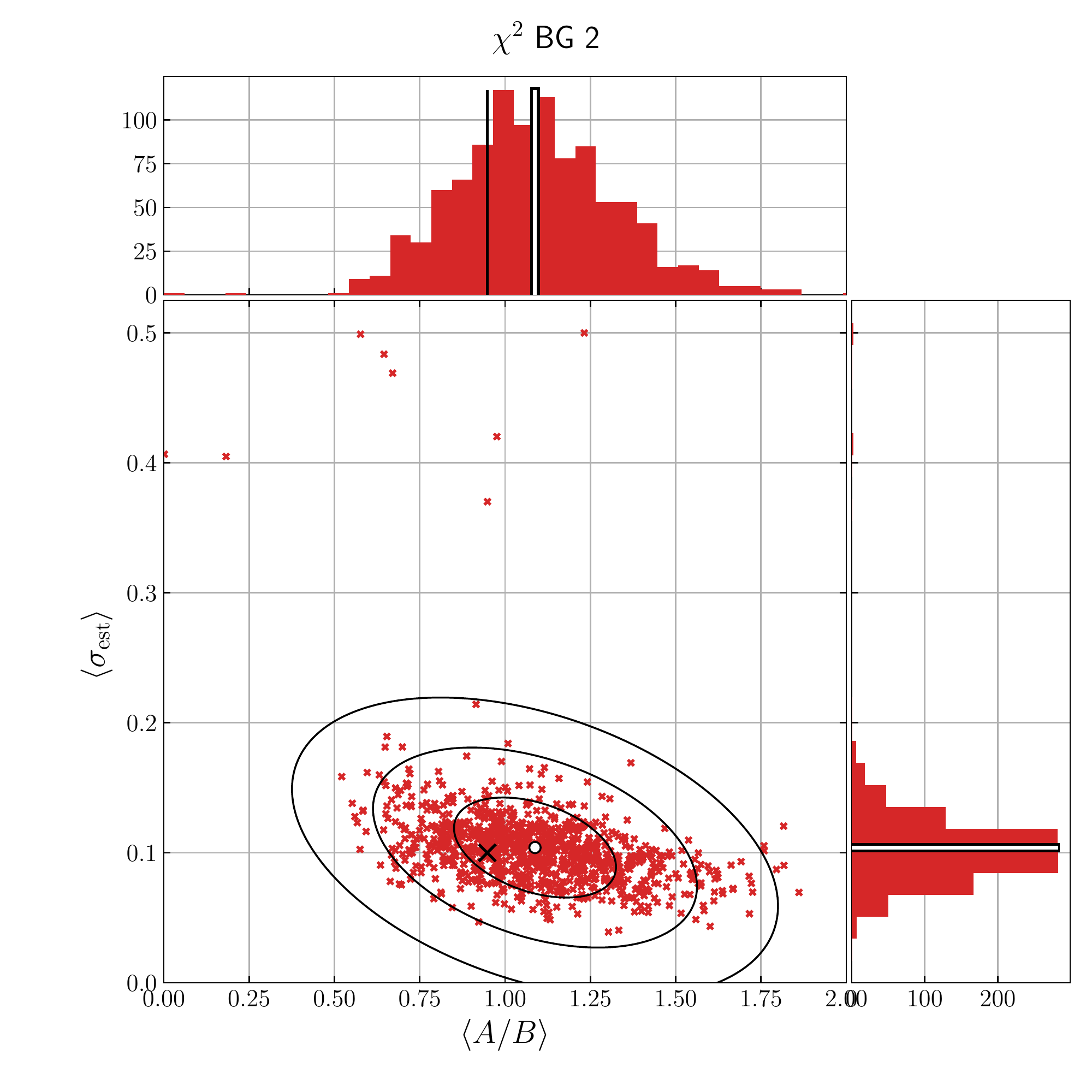}
    \caption{Scatter plot of parameters determined from a series of synthetic data sets, using the Multinomial, Poisson and $\chi^2$ for error scheme 1 and 2 methods, plotted together with histograms of distribution of the fitted value background and amplitude, their true parameters being $A/B$ = 0.926 and $\sigma$ = 0.1. The 1, 2, and 3 $\sigma$ co-variances are plotted as ellipses,  the true and average fitted parameter. Mean value is signified by an empty circle while the true value is signified by a cross.}
    \label{fig:ErrorEstimateHistogram}
\end{figure}

\section{Example: Fitting normalized, low-count neutron scattering data}
Our exploration of synthetic data from simple models gave rather clear results in favour of the both the Multinomial and Poisson fitting methods. However, the litmus test would be the influence of the methods on real-world scattering data.

To investigate this, we use an inelastic neutron scattering data set from a measurement of spin waves in MnF$_2$. 
We chose this system as a demonstration case, because MnF$_2$ has simple inelastic features consisting of only one spin wave branch, as well as the fact that a large single crystal of great quality was available. Important in this context is that since MnF$_2$ orders in an antiferromagnetic structure, different parts of the spin wave spectrum have different intensities. In particular, the magnon intensity around the magnetic Bragg peaks with Miller indices $H+K+L$ being odd is high, as opposed to low close to the structural Bragg peaks, for $H+K+L$ even. This allows for a systematic change of peak intensity when performing 1D cuts for constant energy in a given $Q$ direction. 

All data presented were taken during the early commissioning of the new cold-neutron multiplexing spectrometer CAMEA (PSI) in November-December 2018 \cite{Groitl2016,LassCAMEA2020}.
We used a 6.2~g single crystal sample, held at a temperature of 2~K. The measurements were taken as pure sample rotation scans, using two settings of the analyzer-detector tank and four values of the incoming energy, $E_i$. All conversions, normalizations, and visualizations are performed by the novel MJOLNIR software \cite{LassMJOLNIR2020} developed especially for data from CAMEA-type spectrometers. Because of the binning applied to the data, measurement points close to each other in reciprocal space are added together and thus, a completely true normalization is impossible. Instead, the nomalization is found as the average of normalizations from each detected point being binned into the specific pixel. That is, if 15 different detector pixels are binned together into a single point, the normalization is an average of these.

\begin{figure}[ht]
    \centering
    \includegraphics[height=0.33\linewidth]{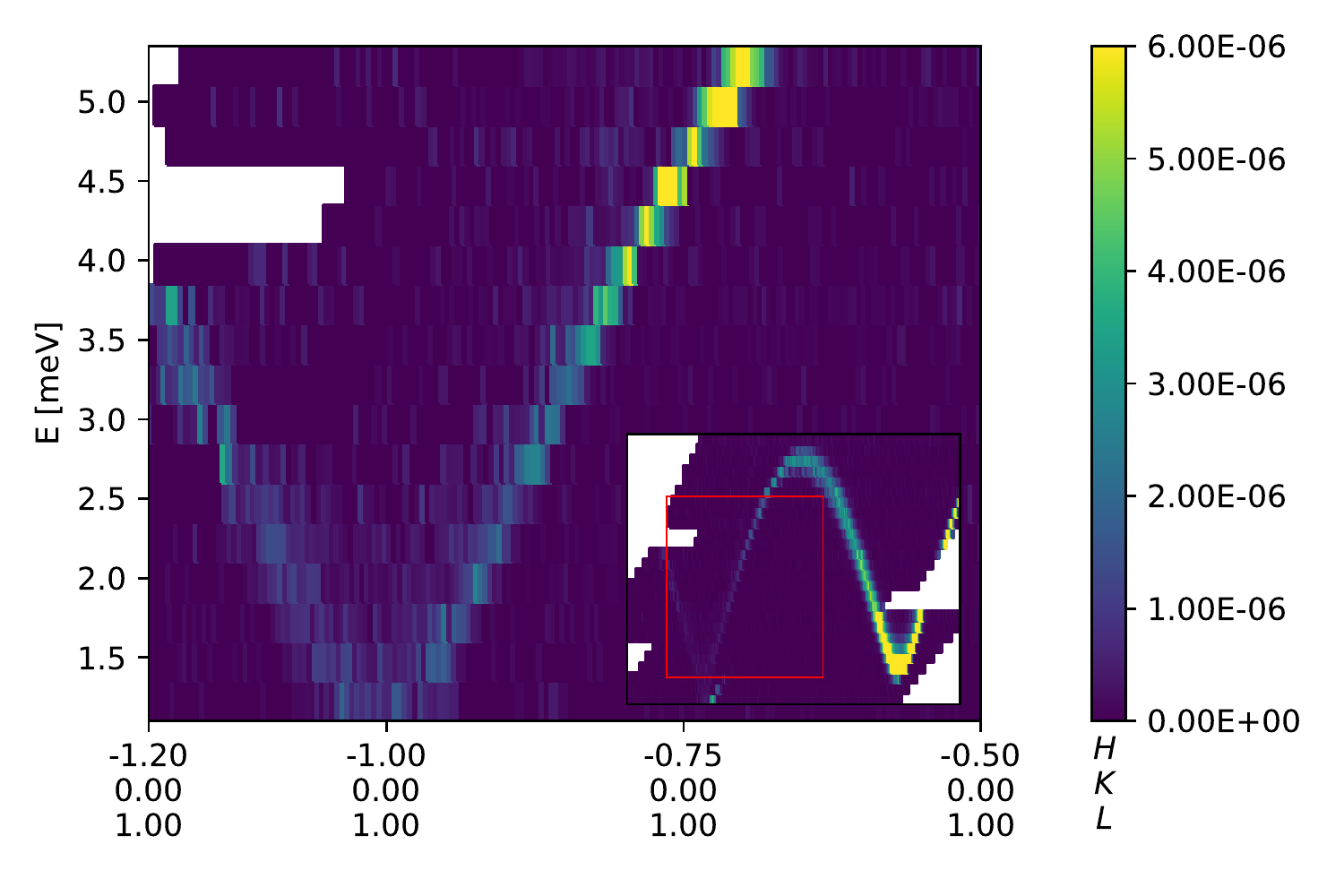}
    \includegraphics[height=0.33\linewidth]{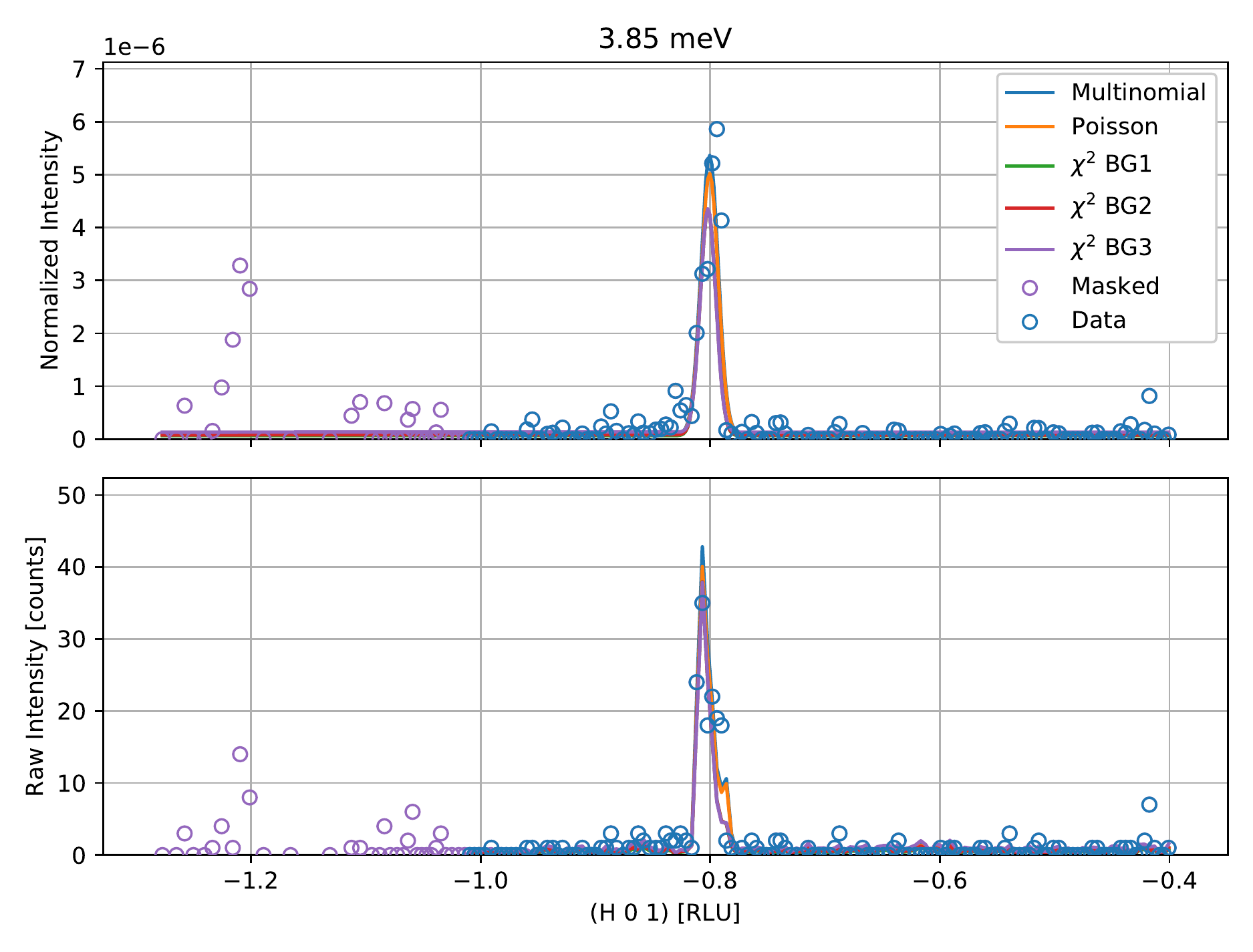}
    \caption{\textbf{Left}: Cut along (H\; 0\; 1) for MnF$_2$ as measured at CAMEA. \textbf{Insert}: Show of full dispersion with color scale a factor of ten larger. \textbf{Right}: excerpt of cut at 3.85 meV with a width of 0.15 meV with corresponding fits. \textbf{Top}: Normalized intensity where sensitivity and monitor/counting time is included. \textbf{Bottom}: Raw neutron counts.}
    \label{fig:MnF2Overview}
\end{figure}

The main data is shown as a color plot in Fig.~\ref{fig:MnF2Overview} (left). A full view of all of the cuts is found in Appendix~\ref{Appendix:FullSet}. We observe a smooth and sharp spin wave dispersion with maximum intensity at the single ion anisotropy gap at 1.0~meV at (0\; 0\; 1) and vanishing intensity close to (-1\; 0\; 1), as known from earlier studies \cite{Yamani2010}. 
We analyse the data by one-dimensional constant-$E$ cuts along $(h 0 1)$, as shown in Fig.~\ref{fig:MnF2Overview} (right).
Analyzing the cuts for different energies shows that the intensity increases with distance to the structural Bragg peak (-1\; 0\; 1). 

Comparing the raw counts with the normalized intensity (Fig.~\ref{fig:MnF2Overview} (right)) illustrates that one cannot use the (unnormalized) raw counts directly to fit the data. Instead a model is imposed on the data consisting of three parts: First the peak model, $\lambda_i^*$, which is here assumed to be a Gaussian peak on a constant background described by eq.\eqref{eq:GaussianPeak}. Second, the sensitivity of the individual analyser-detector pairs, measured by vanadium scattering, and here denoted the \textbf{Normalization}, $N$. Lastly, the dependence on counting time through the \textbf{Monitor} value, $M$. 

In total the model reads 
\begin{equation}
    \lambda_i = \lambda_i^* N_i M_i .
\end{equation}

This combined model has been fitted to the raw neutron counts using the fitting methods discussed above using \texttt{Minuit} and their errors are found using the \texttt{Minos} algorithm. For the human eye, the validity of the fit is very hard to evaluate, see the bottom right of Fig.~\ref{fig:MnF2Overview}, because of the the erratic nature of the normalization, $N_i$. It is much better to look at the model, $\lambda_i$ with the data normalized, top right of the figure. However, to the minimization algorithm, this oddly looking model is straight forward to evaluate.

\begin{figure}[!ht]
    \centering
    \includegraphics[width=0.45\linewidth]{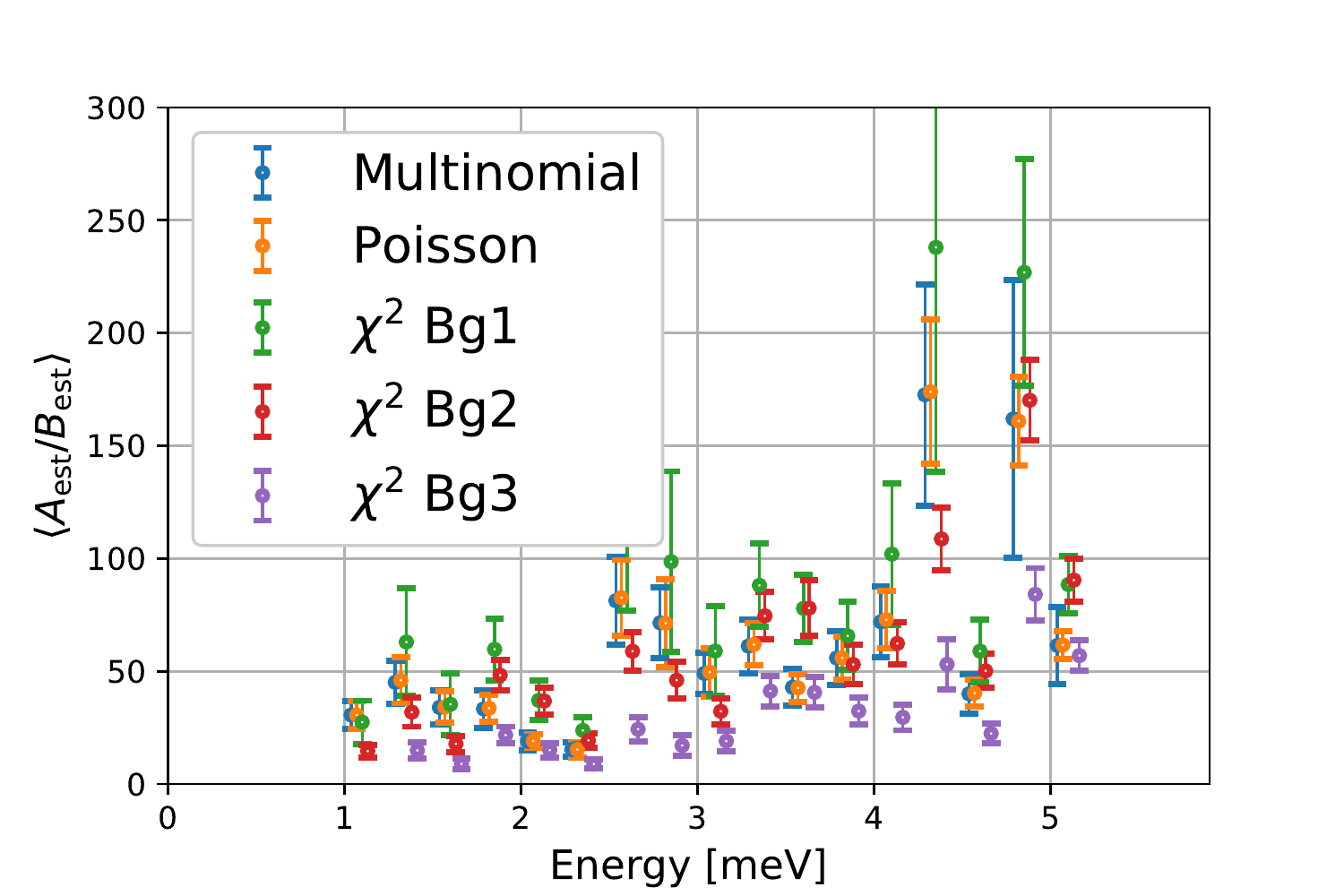}
    \includegraphics[width=0.45\linewidth]{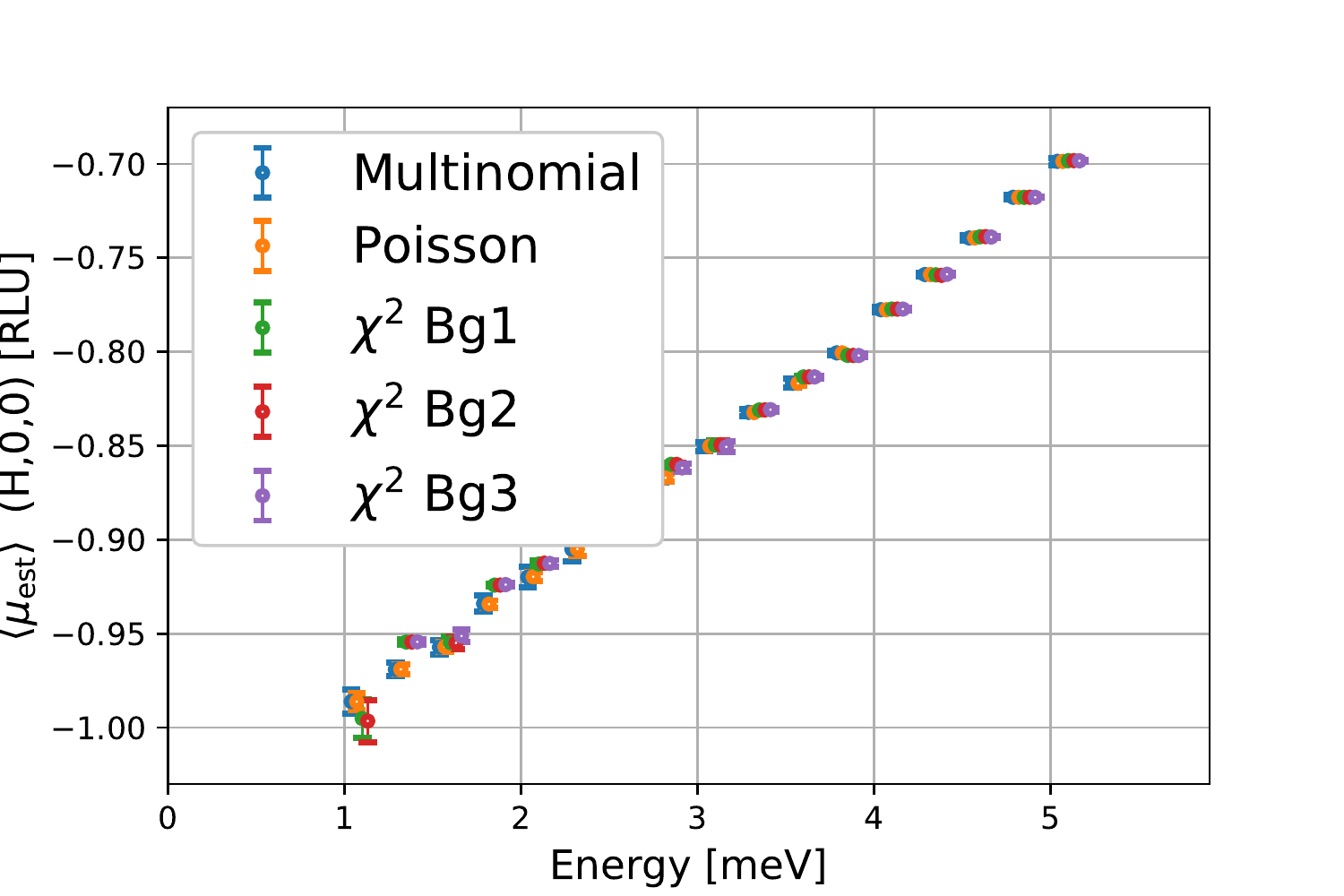}
    \includegraphics[width=0.45\linewidth]{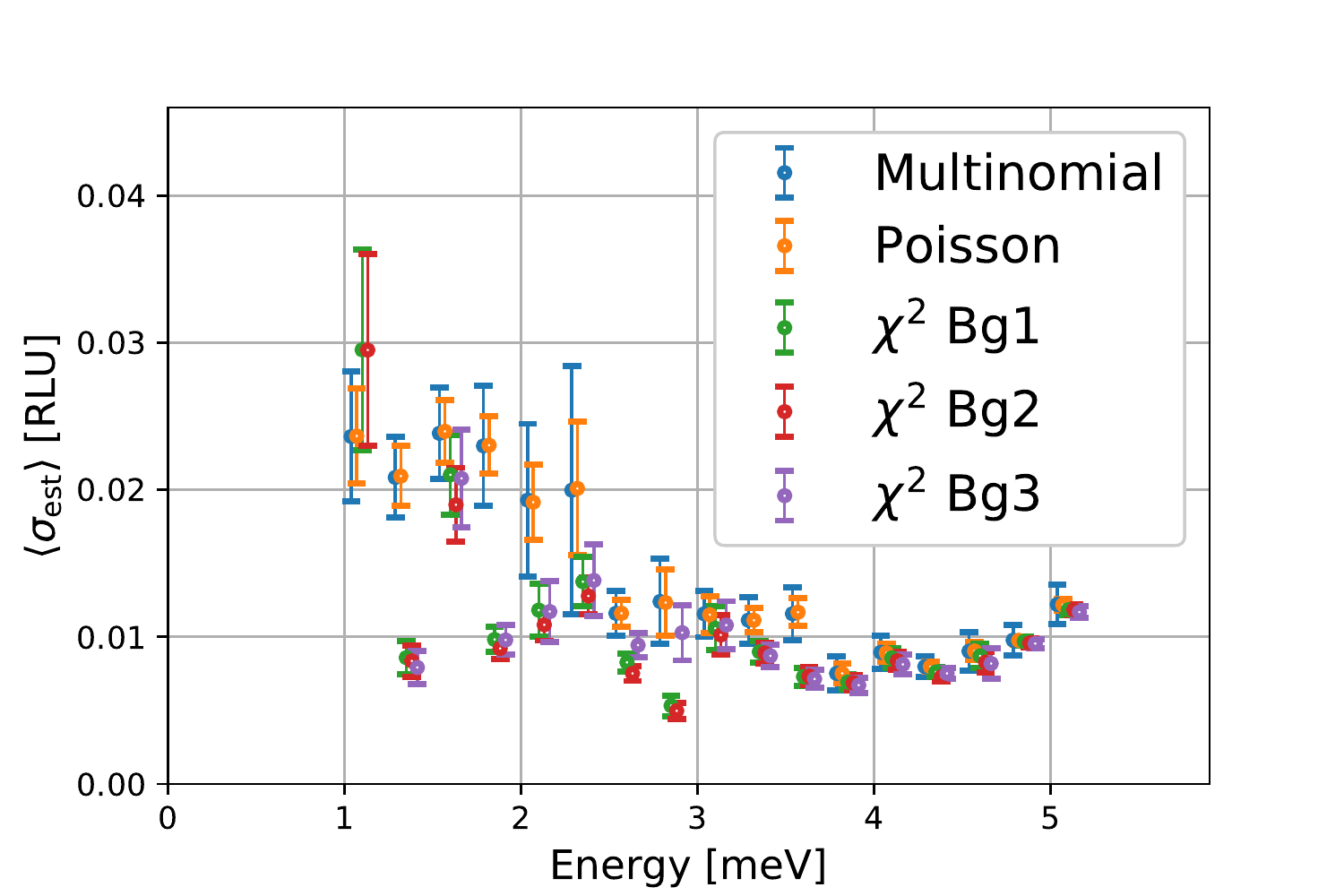}
    \caption{Parameter values and uncertainty as fitted to the MnF$_2$ data using the 5 different methods and \texttt{Minuit}.}
    \label{fig:MnF2FitParameters}
\end{figure}

Fig.~\ref{fig:MnF2FitParameters} shows the outcome of the data analysis for the four different fitting methods.
From the found parameters it can be seen that all methods agree largely on the determination of $\mu$, while some spread is present for the other parameters. As a trend, the Poisson method predicts the largest values for the width of the peak. All methods agree that the amplitude of the peak grows for larger energies, but the exact value is disputed. As the background level in general is low, it is expected that the least squares method where zero counts are excluded overestimates the background and in turn underestimates the amplitude to background. This is indeed the case and this estimate lies below. The Poisson and Multinomial fits agree across all three parameters and at all energies. In order for the fits to converge properly, special care had to be given not to include any intensity to the data that was not described by the fitting function. Especially the background estimation of the Poisson and Multinomial fitting were sensitive to any other structures in the data, see appendix \ref{sec:PoissonWrong}. Looking at the estimated uncertainties in the parameters, it is seen that the Multinomial distribution provides the largest error estimates as compared to the Poisson. This was also seen in Ref.~\ref{sec:Uncertainty} where the conclusion is that the error estimates might be a little to large when compared to the standard deviation on a large sample of spectra.

This example of inelastic scattering data has been measured using a multi-detector setup but with a scan over sample rotation. This results in the requirements for using the Multinomial not being met, but rather one is to use the Poisson formalism. 

\section{Discussion and conclusion}

Three different log-likelihoods have been presented having their origin in the Multinomial, Poisson and Gaussian distributions. We have reviewed the formalism to perform data analysis through parameter estimation using these in order to tackle the difficulties arising when performing simple $\chi^2$ fits on counting data. 

When treating a featureless spectrum we have clarified analytically and by the use of synthetic data that the Gaussian approximation of the Poisson distribution, both inside and outside the Poisson regime, will result in a clear bias. The simple reason is that Gaussian statistics weigh low-count data points higher. 
In contrast, the Poisson fits yields unbiased results, while the Multinomial method simply (and correctly) provides the mean count value as a constant value.


When fitting synthetic data with a peak on a constant background it was shown that the Multinomial and Poisson fitting methods produced much better results for parameters as compared to the three different least squares method. All of the methods had, on average, a good estimation of the peak center, with the Multinomial having the smallest standard deviation. In addition, in both the signal-to-noise and width parameters the $\chi^2$ methods were biased. When the signal-to-noise level decreased below 1, also the Poisson method became biased. A small bias was also found for the Multinomial in $A/B$, but was found to be reduced by increasing the number of counts in the spectrum. By investigating a single spectrum, it was found that a cross correlation between peak width and signal-to-background level was present. It was further found that the loglikelihood of the Multinomial distribution increased the slowest signifying a larger area of uncertainty. This is rather artificial as the signal-to-background parameter for all but the Multinomial distribution is a combination of two fitted parameters, where the background uncertainty was not propagated. Further, the spread of multiple estimated parameters for many spectra yield a smaller standard deviation for Multinomial and Poisson as compared to the Gaussian methods.

One of the drawbacks of using Multinomial and Poisson statistics is the need for maintaining the original count values of all data, for example in case of efficiency and monitor normalization as well as background subtraction. This complication makes development of data analysis software one level more complex. Nevertheless, we have implemented such a framework in the MJOLNIR analysis package and used this to compare the Poisson and Gaussian methods on simple, but real, data on spin waves in MnF$_2$.
Our findings show that the Multinomial and Poisson methods are less stable than the Gaussian methods with regards to the case where the fitting function does not fully describe the data. This necessitated masking away data regions containing other signals than the peak being fitted. For all of the Gaussian methods, such a masking procedure was not necessary in order to get an acceptable fitting result.

It is only in the case of a one-shot acquisition that the Multinomial distribution is correct, i.e. when all data points are measured at the same time. If a scan is performed it is actually the Poisson distribution that is to be used. Despite the Multinomial and Poisson statistics being the correct methods in each their setting, the sturdiness and reliability of the Gaussian least-square certainly counts in the favour of this well-established method. 


In conclusion, the Multinomial, Poisson and Gaussian methods have their strengths and justification, and we advocate that future full-fetched analysis programs should be equipped with more than one fitting method for their data analysis algorithms. A process could consist of first a $\chi^2$ fitting optimizing user provided initial guesses followed by a log-likelihood fit using Poisson or Multinomial statistics as needed.

\section*{Acknowledgements}
It is a pleasure to thank Toby Perring, Tobias Weber and Dmitry Gorkov for valuable discussions regarding understanding of error estimates and visualization of model and parameter uncertainties. We would also like to thank Christof Niedermayer for the assistance while obtaining the MnF$_2$ data set at CAMEA.

\newpage
\section*{Appendix}

\renewcommand{\thesubsection}{\Alph{subsection}}

\subsection{Multinomial Log-likelihood derivation}\label{app:Multinomial}
Taking the logarithm, and applying Stirling's approximation for all faculty terms, one gets
\begin{align}
\ln{L} \approx& N\ln{N}-N \sum_{i=1}^m \sbrac{n_i\ln{p_i}-n_i\ln{n_i}+n_i}+\parh{N-\Delta N}\ln{1-\Delta P}\notag \\&-\parh{N-\Delta N}\ln{N-\Delta N}+\parh{N-\Delta N}\\
=&N\ln{N}+\parh{N-\Delta N}-\Delta N-\parh{N-\Delta N}\ln{N-\Delta N}+\sum_{i=1}^m \sbrac{n_i+n_i\ln{\frac{p_i}{n_i}}}\notag \\&+\parh{N-\Delta N}\ln{1-\Delta p}\\
=&N\ln{N}+\sum_{i=1}^m\sbrac{n_i\ln{\frac{p_i}{n_i}}}+\parh{N-\Delta N}\ln{\frac{1-\Delta p}{N-\Delta N}}\\
=&\sum_{i=1}^m\sbrac{n_i\ln{\frac{p_iN}{n_i}}}+\parh{N-\Delta N}\ln{\frac{1-\Delta p}{1-\frac{\Delta N}{N}}},
\end{align}
where the last equality follows from rewriting $N\ln{N} = \parh{N-\Delta N}\ln{N}+ \sum_{i=1}^m n_i\ln{N}$. Instead of working directly with the quantities, normalized ones is introduced
\begin{align}
q_i =& \frac{n_i}{\Delta N},\qquad \sum_{i=1}^m q_i = 1 \\
\tilde{p}_i =& \frac{p_i}{\Delta p}, \qquad \sum_{i=1}^m \tilde{p}_i = 1 \\
\end{align}
Inserting these into the log-likelihood, 
\begin{align}
\ln{L} = & N\sbrac{\frac{\Delta N}{N}\sum_{i=1}^m\parh{\frac{n_i}{\Delta N}\ln{\frac{p_i}{\Delta p}\frac{\Delta p}{\frac{n_i}{\Delta N}}\frac{N}{\Delta N}}}+\parh{1-\frac{\Delta N}{N}}\ln{\frac{1-\Delta p}{1-\frac{\Delta N}{N}}}}.
\end{align}
Reducing this one finally reaches
\begin{equation}
    \ln{L}= N\sbrac{\frac{\Delta N}{N}\sum_{i=1}^m\parh{q_i\ln{\frac{\tilde{p}_i}{q_i}}}+\frac{\Delta N}{N}\ln{\frac{\Delta p}{\frac{\Delta N}{N}}}+\parh{1-\frac{\Delta N}{N}}\ln{\frac{1-\Delta p}{1-\frac{\Delta N}{N}}}}.
\end{equation}

\begin{align}
\pdx{x_\alpha}\ln{L} = \pdx[\Delta p]{x_\alpha}\pdx[\ln{L}]{\Delta p}+\sum_{i=1}^m\sbrac{\pdx[\tilde{p}_i]{x_\alpha}\pdx[\ln{L}]{\tilde{p}_i}} = 0\label{eq:MaximizeL}
\end{align}
Looking at the derivative of $\ln{L}$ with respect to $\Delta p$,
\begin{equation}
\pdx[\ln{L}]{\Delta p} = N\parh{\frac{\frac{\Delta N}{N}}{\Delta p}-\frac{1-\frac{\Delta N}{N}}{1-\Delta p}}.\label{eq:DiffDeltaP}
\end{equation}
The term inside the summation in eq.~\eqref{eq:MaximizeL} is equivalent to
\begin{align}
\sum_{i=1}^m\sbrac{\pdx[\tilde{p}_i]{x_\alpha}\pdx[\ln{L}]{\tilde{p}_i}} = \pdx[\ln{\tilde{L}}]{x_\alpha}\\
\ln{\tilde{L}} = \Delta N \sum_{i=1}^m q_i\ln{\frac{\tilde{p}_i}{q_i}}.
\end{align}
This follows from 
\begin{align}
\pdx[\ln{L}]{\tilde{p}_i} = \Delta \pdx{\tilde{p}_i}q_i\ln{\frac{\tilde{p}_i}{q_i}} = \Delta N\frac{q_i}{\tilde{p}_i}\\
\sum_{i=1}^m\sbrac{\pdx[\tilde{p}_i]{x_\alpha}\Delta N\frac{q_i}{\tilde{p}_i}} = \Delta N\sum_{i=1}^m\sbrac{\frac{q_i}{\tilde{p}_i}\pdx[\tilde{p}_i]{x_\alpha}} = \pdx[\ln{\tilde{L}}]{x_\alpha}.\label{eq:DiffTildeLxAlpha}
\end{align}

Solving the above equation can be split into two; firstly if eq.~\eqref{eq:DiffDeltaP} is zero, the first term in eq.~\eqref{eq:MaximizeL} is independent of $x_\alpha$, which gives
\begin{equation}
0 = N\parh{\frac{\frac{\Delta N}{N}}{\Delta p}-\frac{1-\frac{\Delta N}{N}}{1-\Delta p}} \Rightarrow \Delta p^* = \frac{\Delta N}{N},
\end{equation}
where $\Delta p^*$ is introduced as the optimal parameter. This is a natural result stating that the maximal $\ln{L}$ is when the modeled number of neutrons not hitting the detector coincides with the real world number. Inserting these values of $\Delta p^*$ in $\ln{L}$ yields
\begin{align}
\ln{L}_{\Delta p=\Delta p^*} &= N\sbrac{\frac{\Delta N}{N}\sum_{i=1}^m\sbrac{q_i\ln{\frac{\tilde{p}_i}{q_i}}}+\frac{\Delta N}{N}+\parh{1-\frac{\Delta N}{N}}} \\
&=N+\Delta N \sum_{i=1}^m q_i\ln{\frac{\tilde{p}_i}{q_i}}.
\end{align}
The best fitting parameters are then found by optimizing 
\begin{equation}\label{eq:OptimizationOfFit}
\sum_{i=1}^m q_i\ln{\frac{\tilde{p}_i}{q_i}} = \sum_{i=1}^m q_i\ln{\tilde{p}_i}-\underbrace{\sum_{i=1}^m q_i\ln{q_i}}_{\mathrm{const}},
\end{equation}
where $q_i\ln{q_i}$ is independent of the optimization parameters $x_\alpha$, which only influence $\tilde{p}_i$. Applying this to fitting, one has to minimize
\begin{equation}
-\ln{P(D|M)} = -\Delta N\sum_{i=1}^m q_i\ln{\tilde{p}_i}.
\end{equation}

\subsection{The full MnF$_2$ data set}\label{Appendix:FullSet}
For completeness, Figs.~\ref{fig:FitExamplesMnF2}, \ref{fig:FitExamplesMnF22}, and \ref{fig:FitExamplesMnF23} presents the full suite of fits to the MnF$_2$ data, discussed in the main text.
\begin{figure}[ht!]
    \centering
    \includegraphics[width=0.45\linewidth]{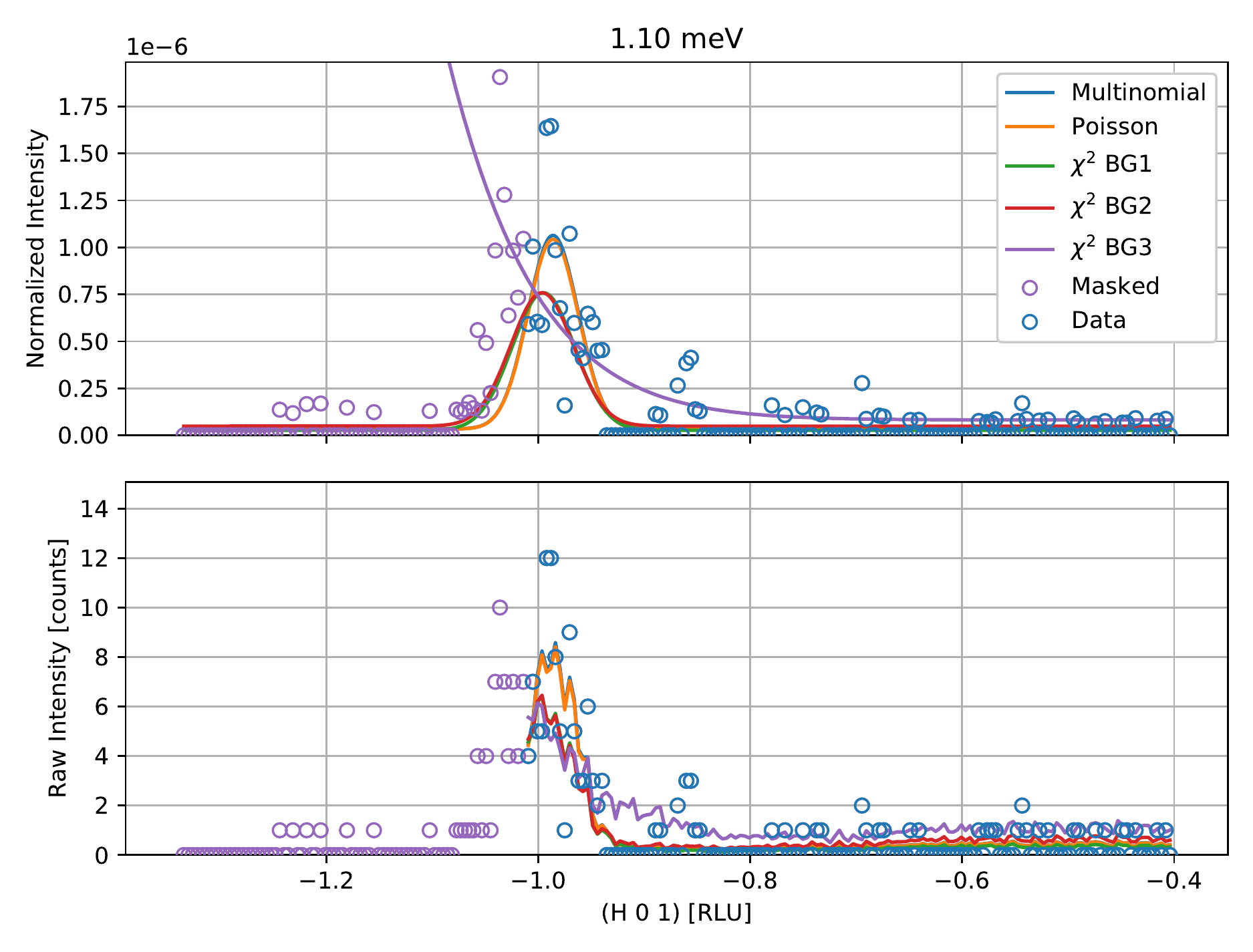}
    \includegraphics[width=0.45\linewidth]{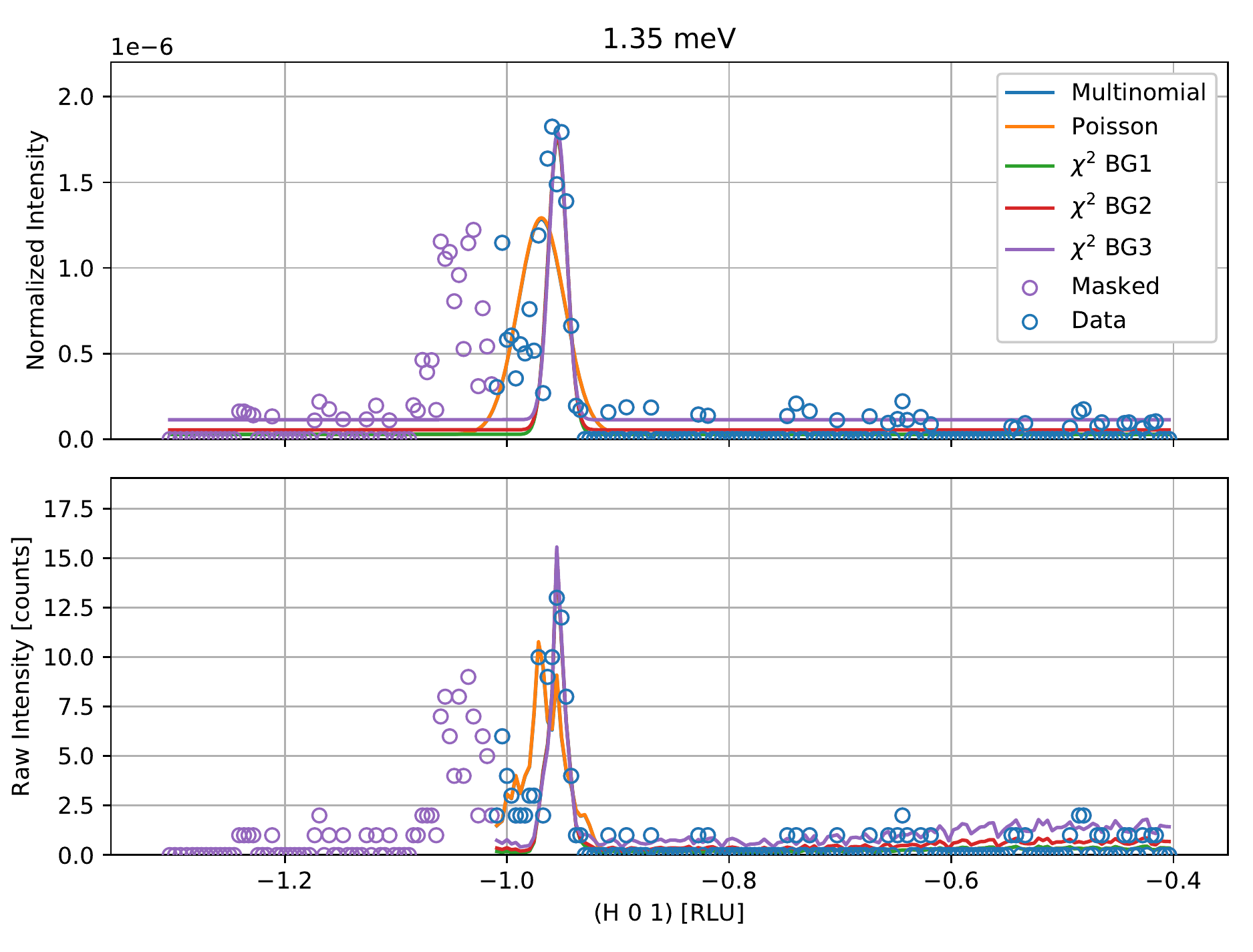}   \includegraphics[width=0.45\linewidth]{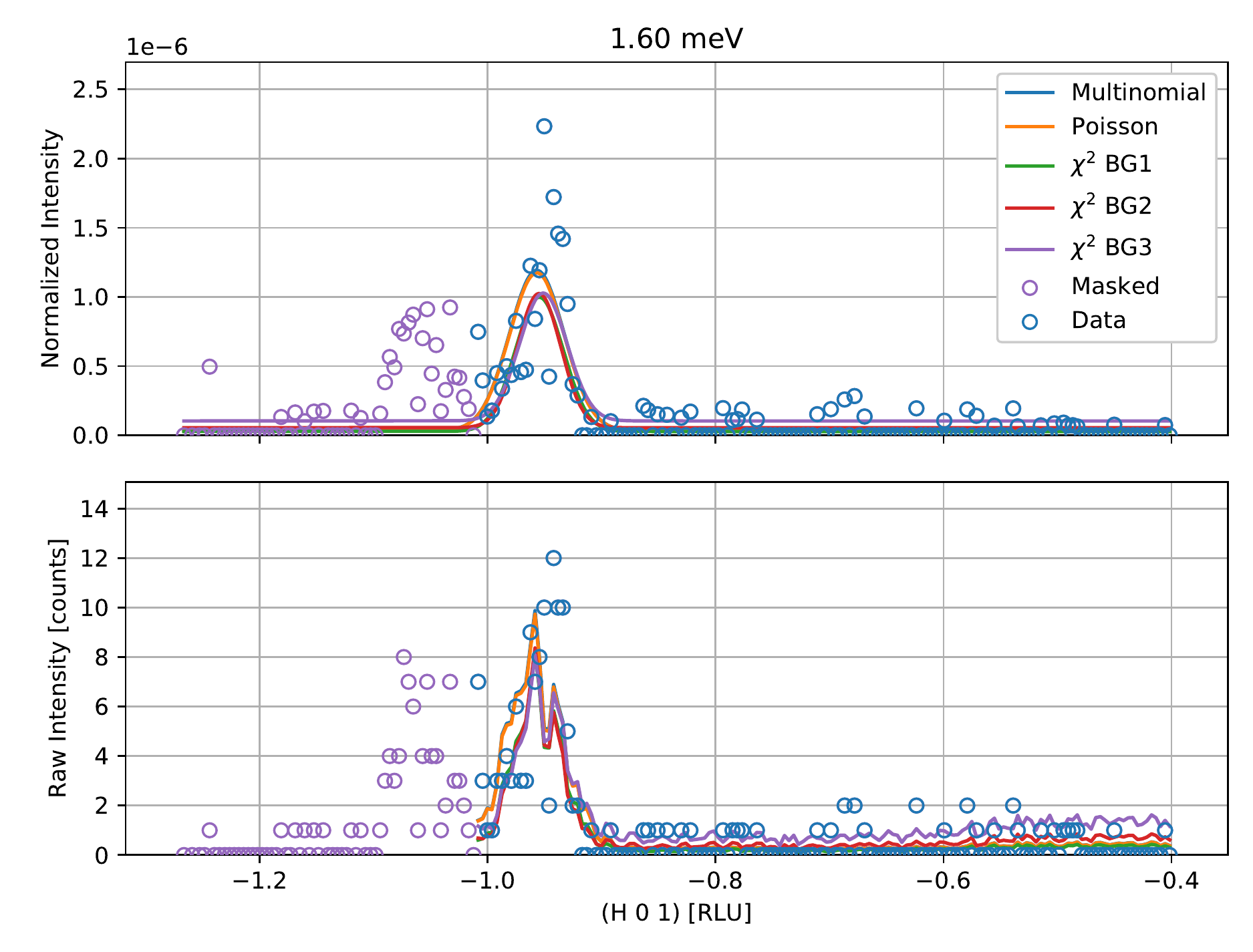}
    \includegraphics[width=0.45\linewidth]{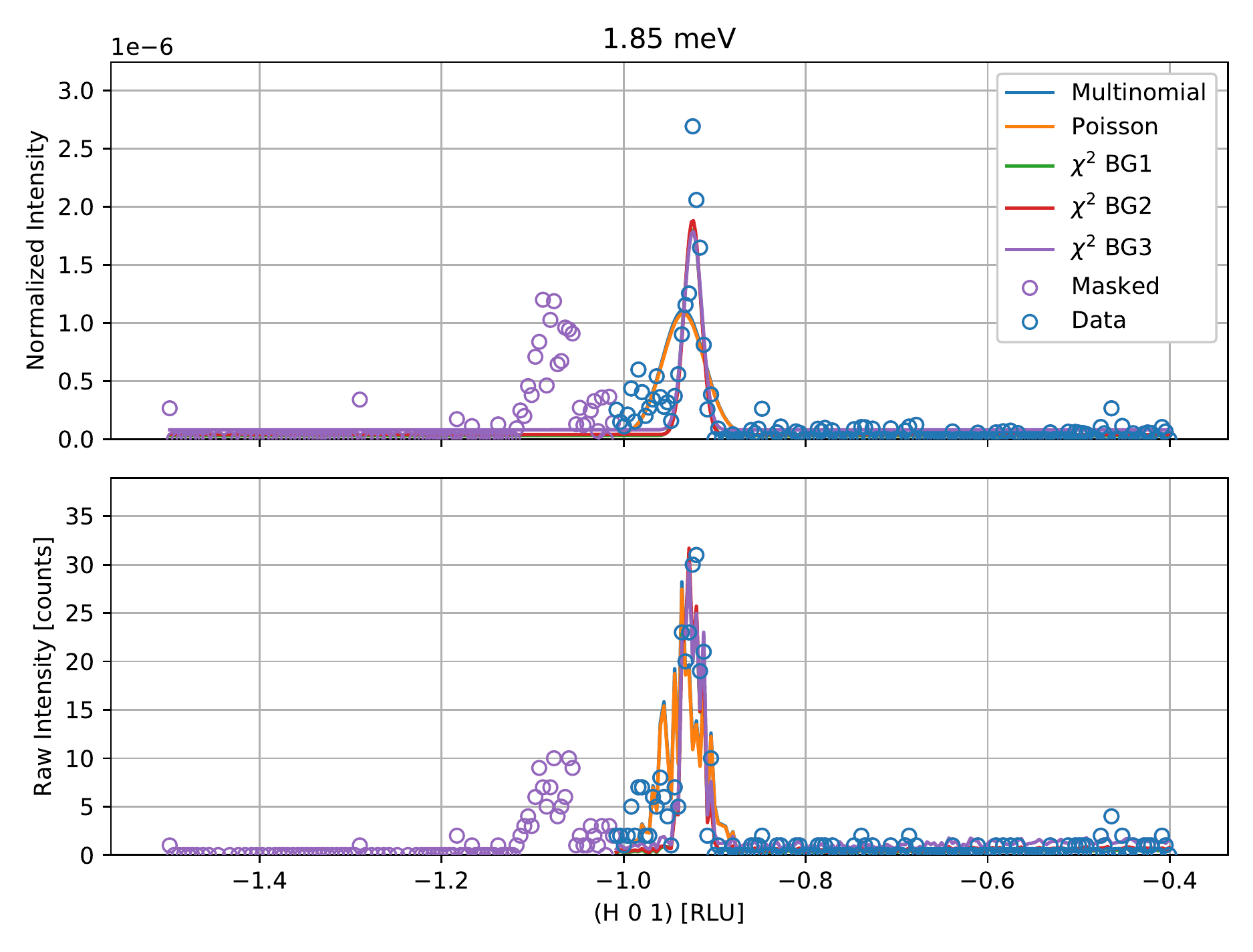}
    \includegraphics[width=0.45\linewidth]{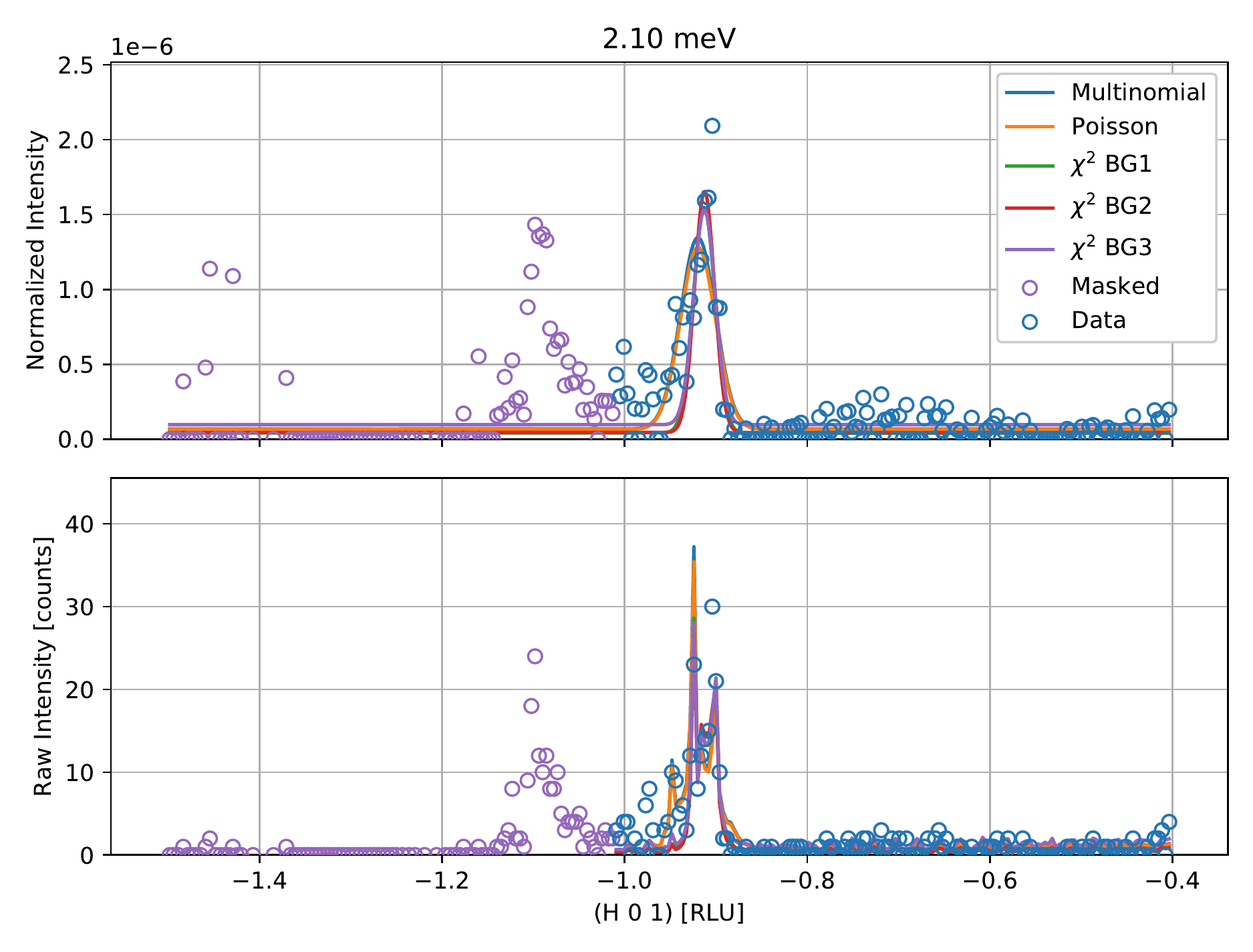}
    \includegraphics[width=0.45\linewidth]{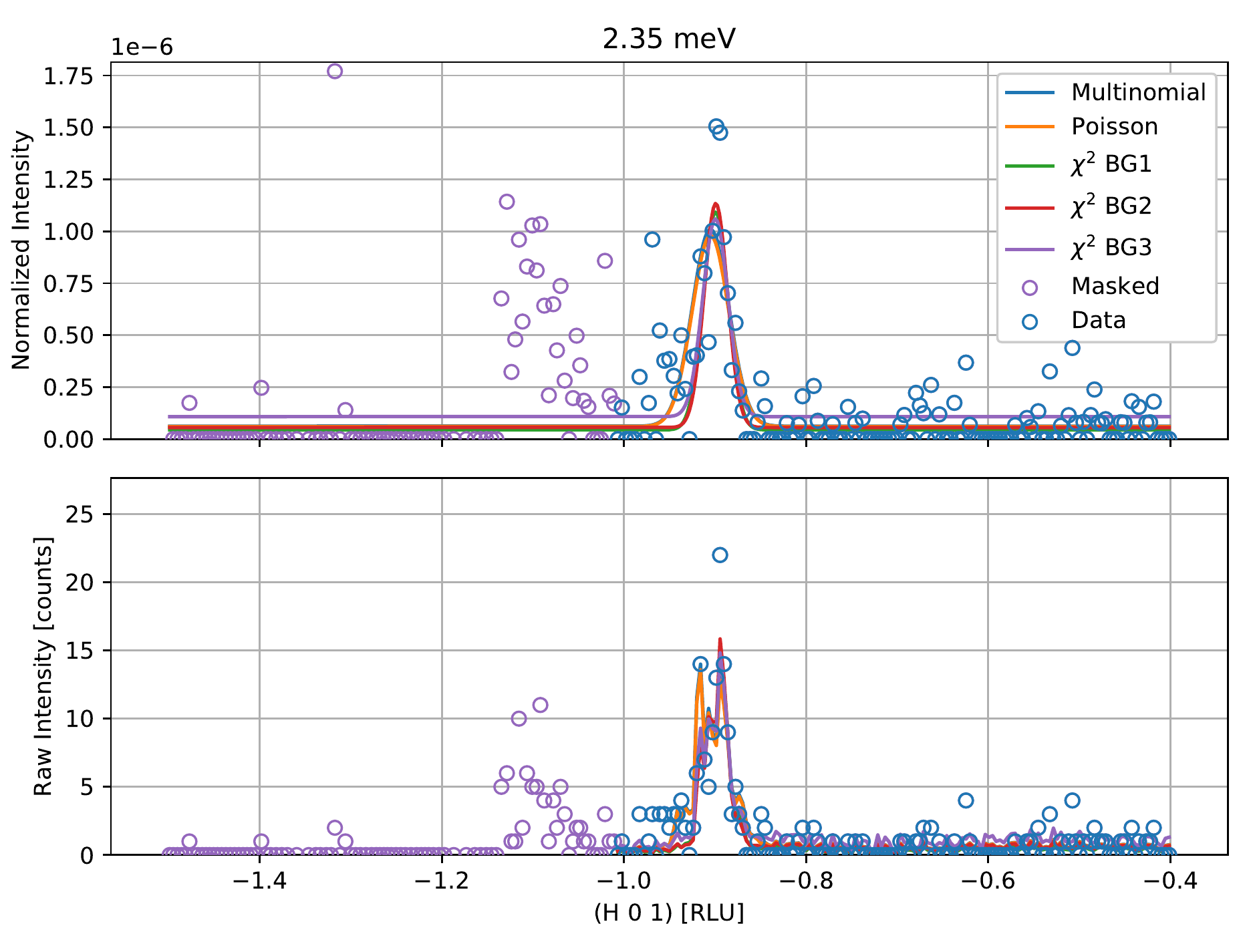}

    \caption{Fit of the Gaussian peak using the Poisson statistics and Gaussian statistic with the three different background methods for the latter. Top panels are normalized data to monitor and normalization while bottom panels are raw counting numbers.}
    \label{fig:FitExamplesMnF2}
\end{figure}
\begin{figure}[ht!]
    \centering
    \includegraphics[width=0.45\linewidth]{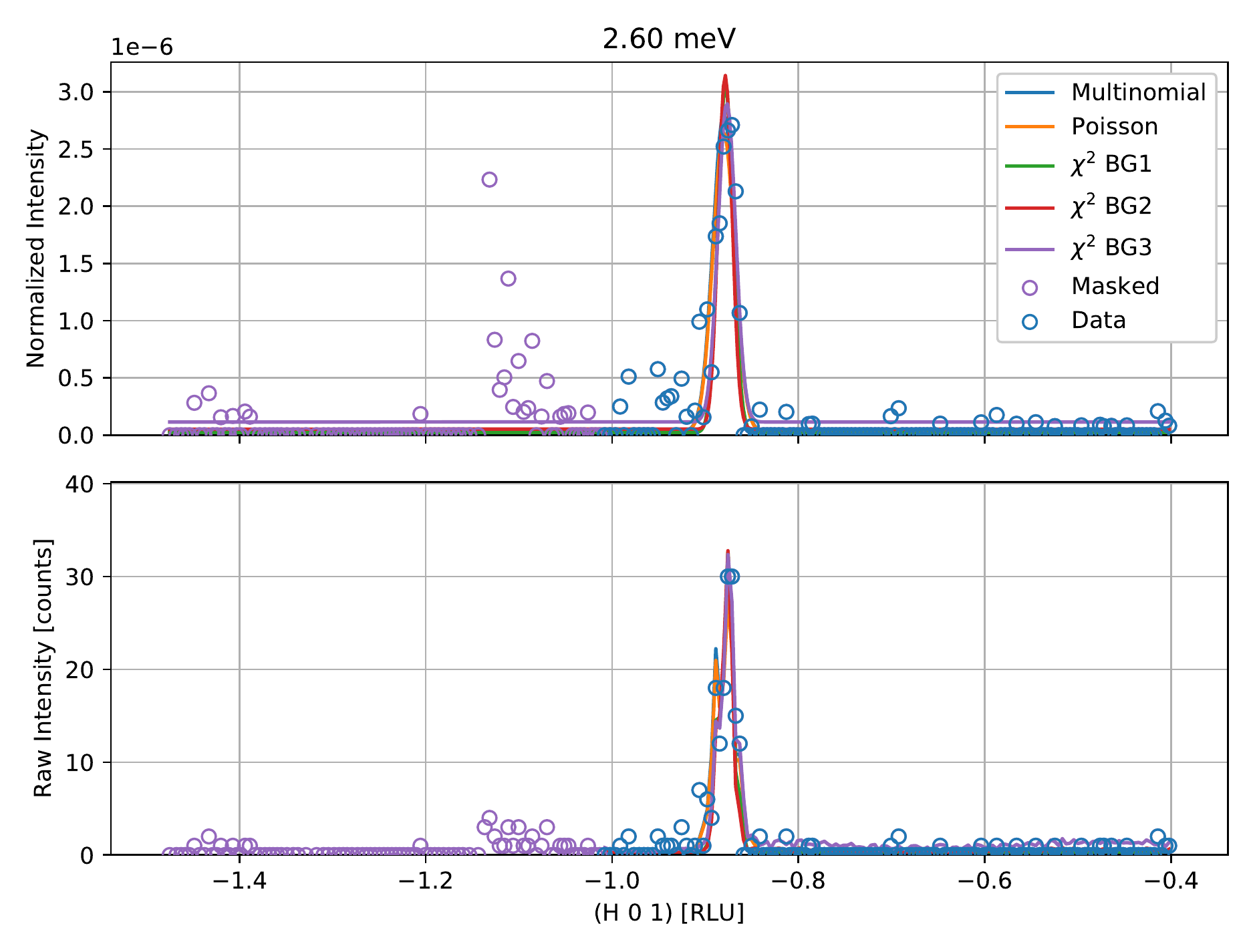}
    \includegraphics[width=0.45\linewidth]{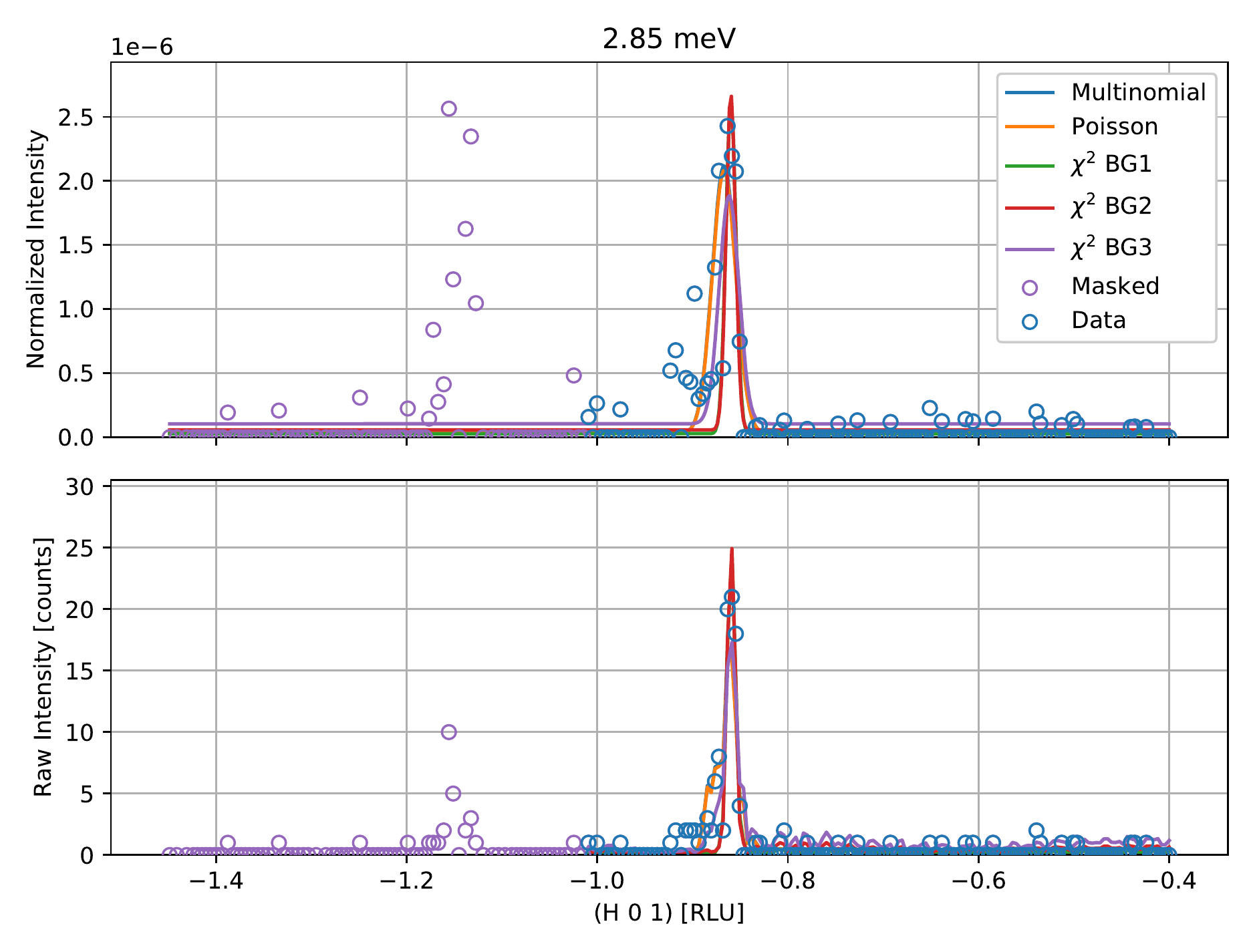}
    \includegraphics[width=0.45\linewidth]{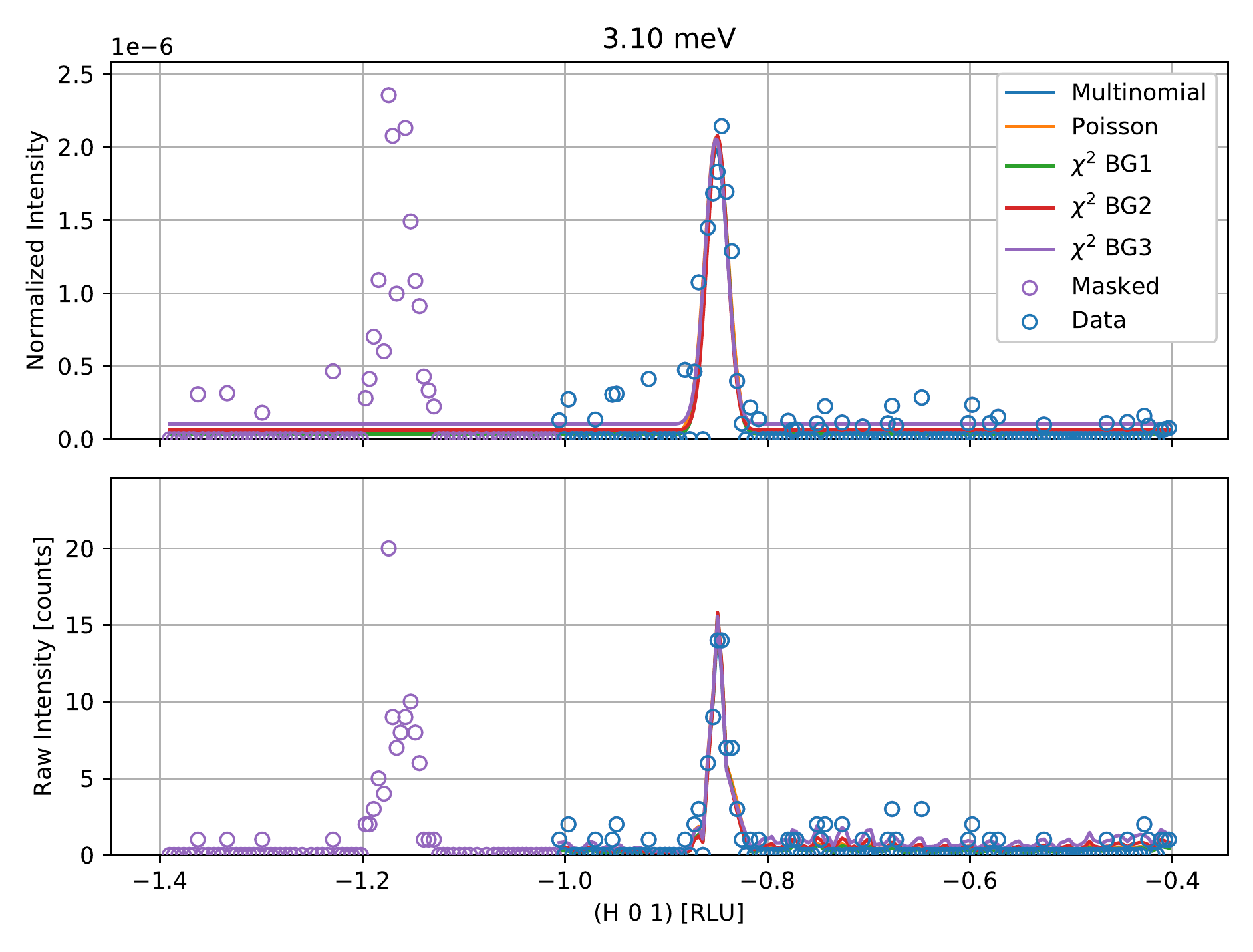}
    \includegraphics[width=0.45\linewidth]{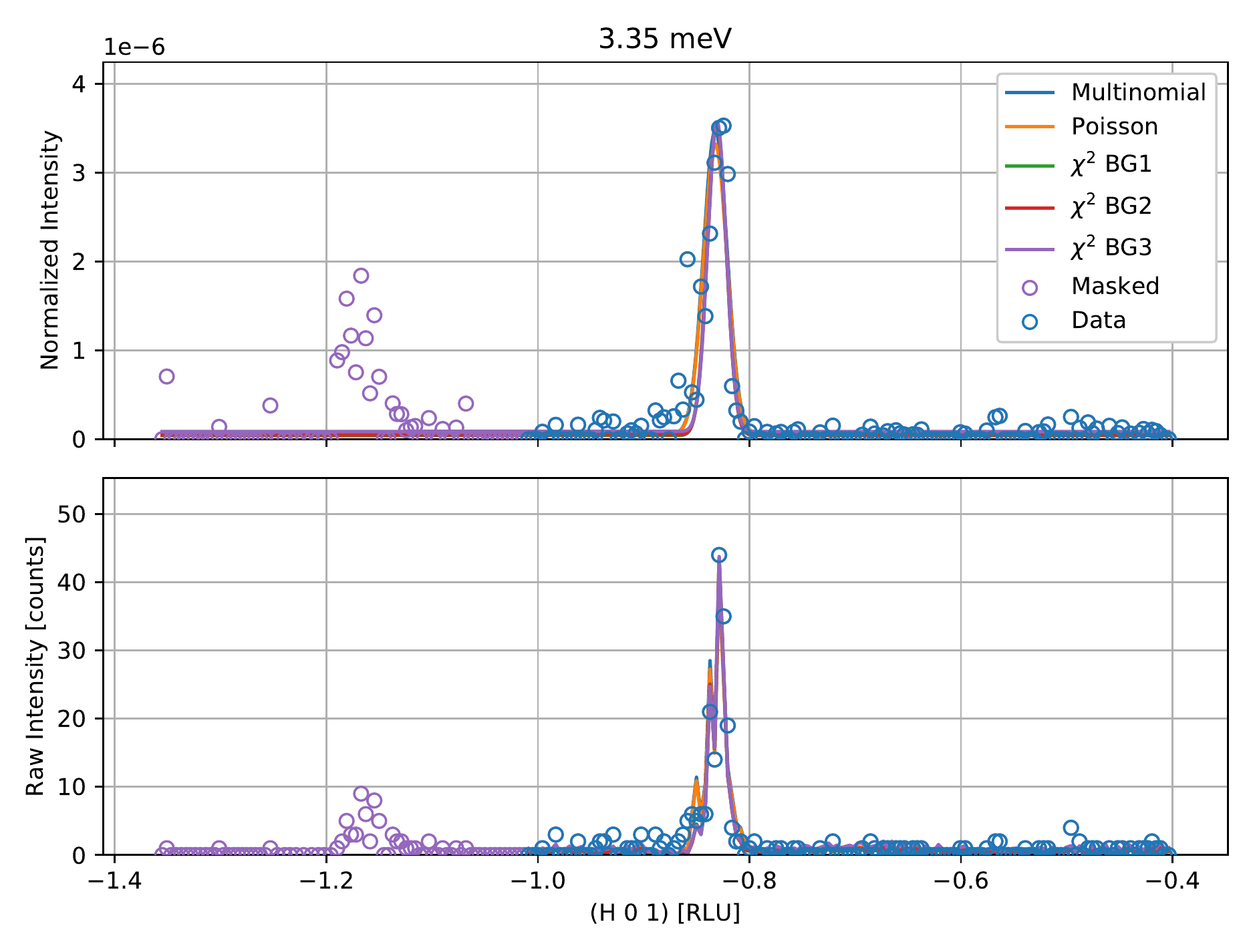}
    \includegraphics[width=0.45\linewidth]{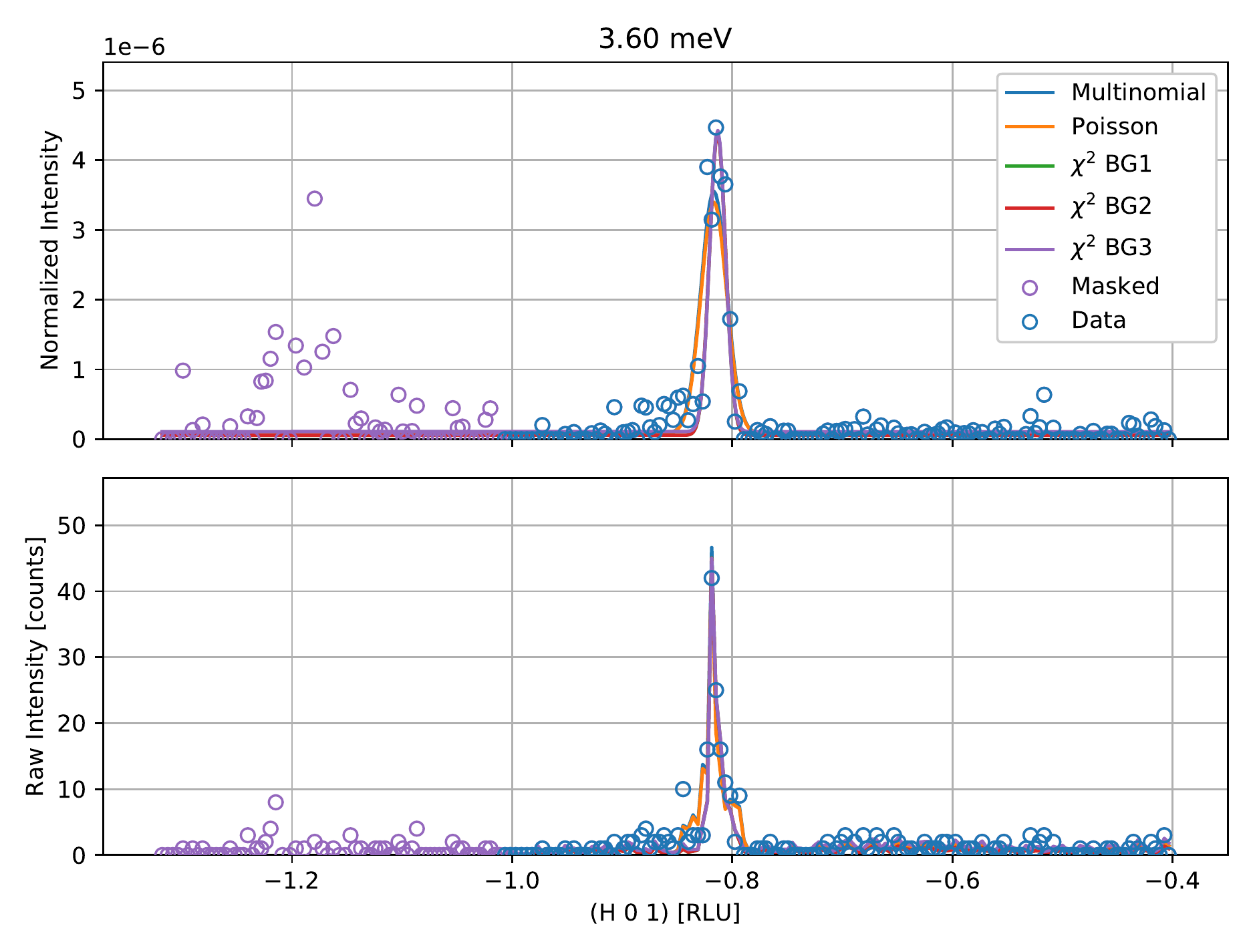}
    \includegraphics[width=0.45\linewidth]{PDFfigures/E3_85Fitted_220420}
    \caption{Continuation of Fig.~\ref{fig:FitExamplesMnF2}.}
    \label{fig:FitExamplesMnF22}
\end{figure}
\begin{figure}[ht!]
    \centering
    \includegraphics[width=0.45\linewidth]{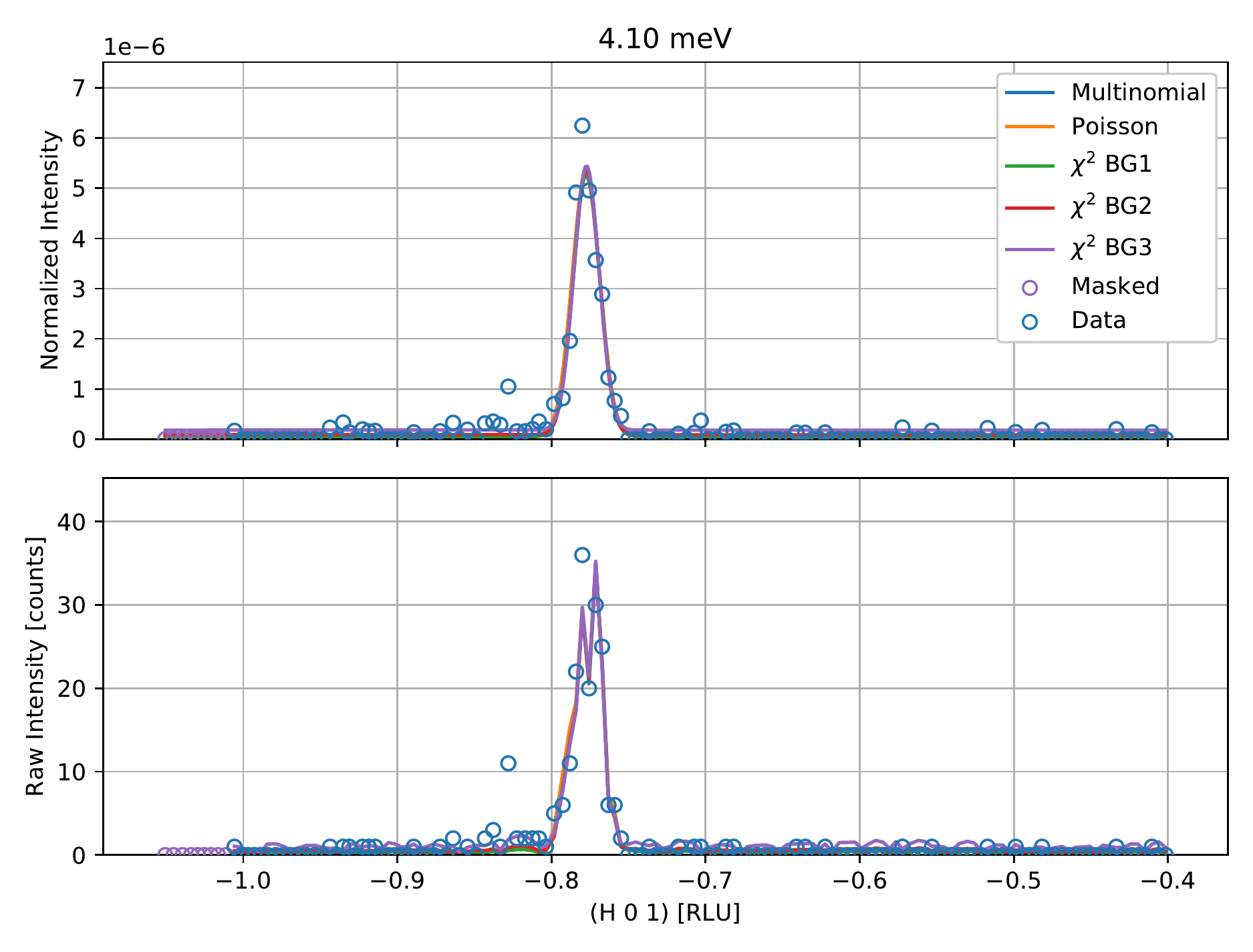}
    \includegraphics[width=0.45\linewidth]{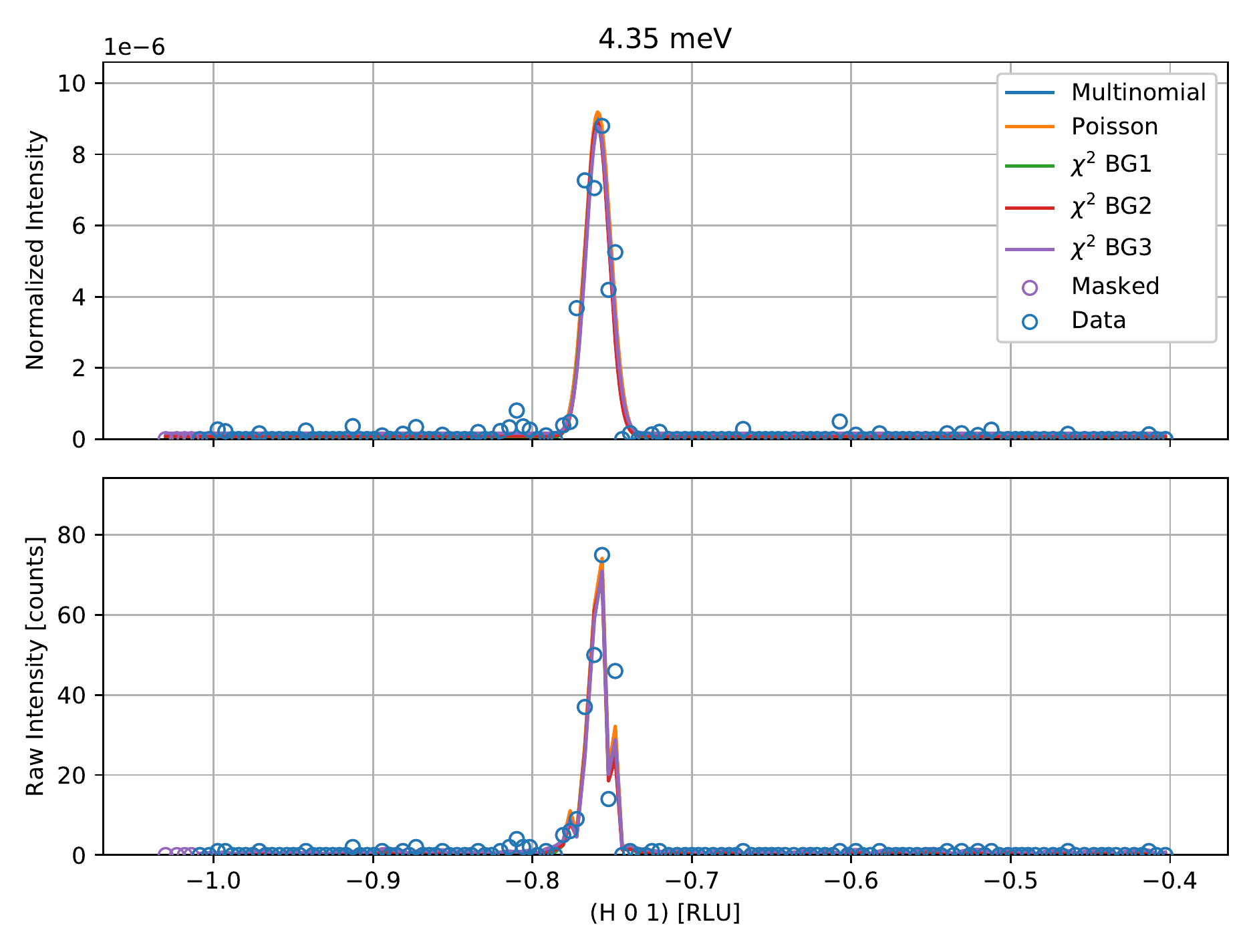}
    \includegraphics[width=0.45\linewidth]{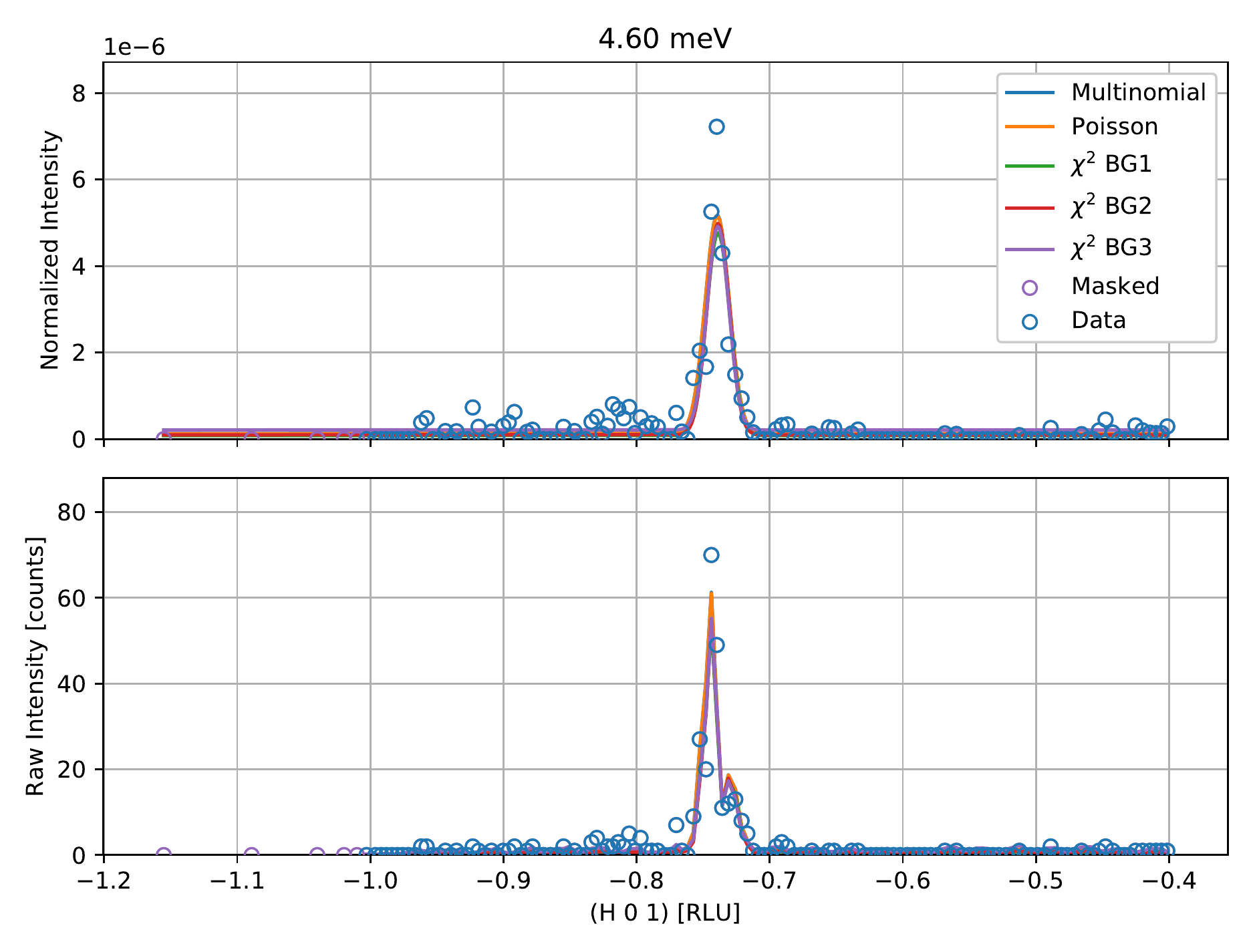}
    \includegraphics[width=0.45\linewidth]{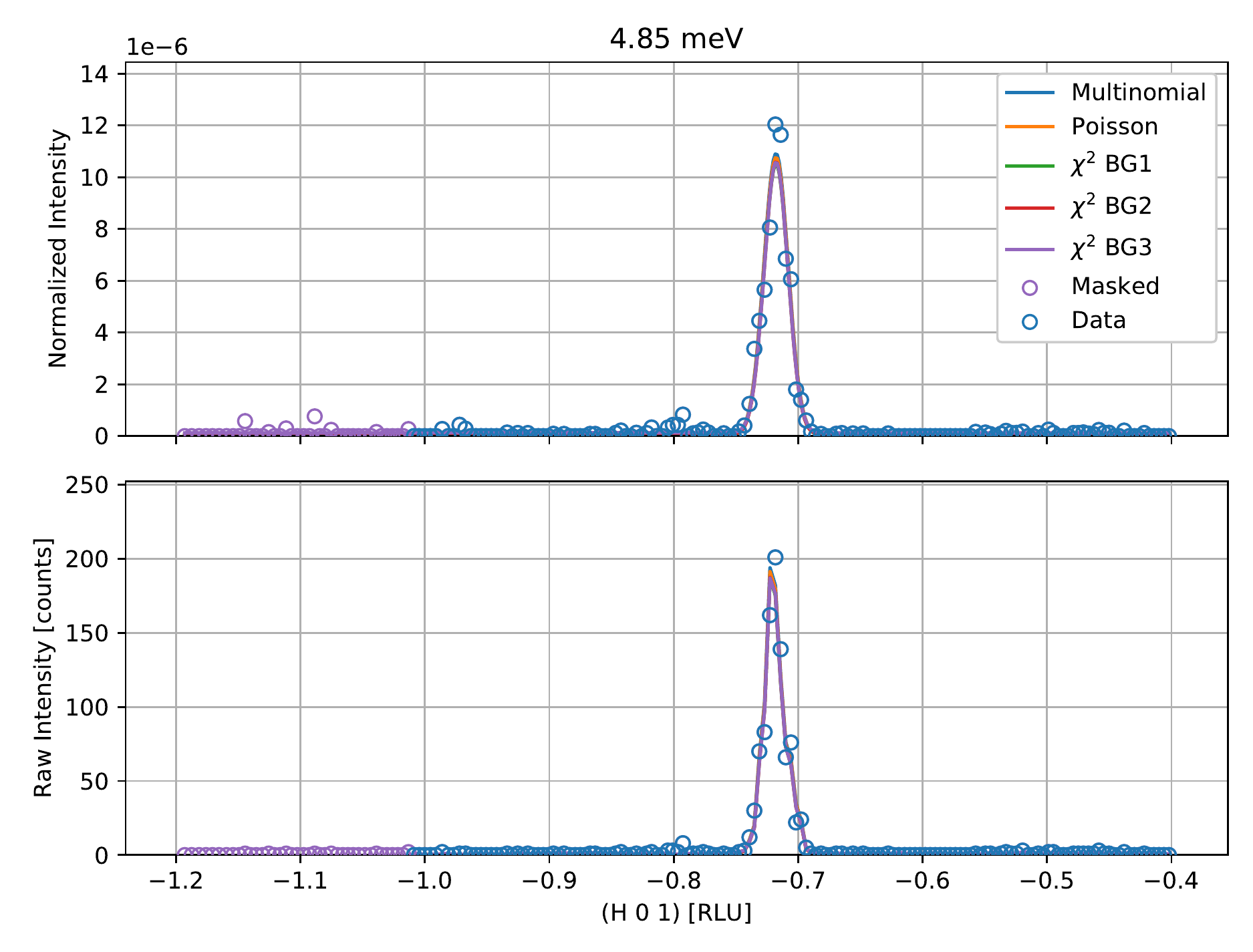}
    \caption{Continuation of Fig.~\ref{fig:FitExamplesMnF22}.}
    \label{fig:FitExamplesMnF23}
\end{figure}

\newpage
\subsection{Proof of the systematic errors in the values of background estimators} \label{app:BackgroundEstimators}
\begin{figure}[h!]
    \centering
    \includegraphics[width=0.45\linewidth]{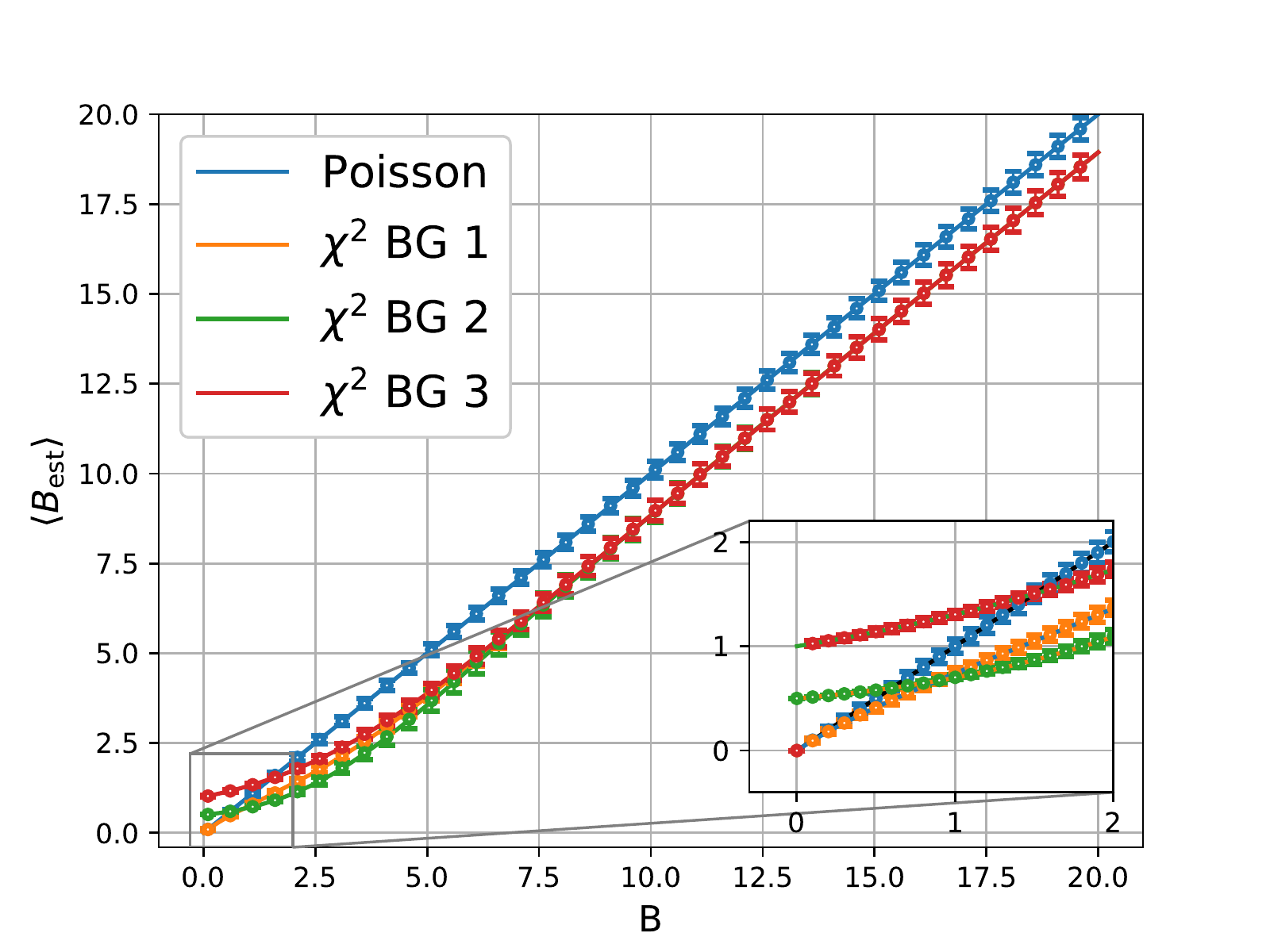}
    \includegraphics[width=0.45\linewidth]{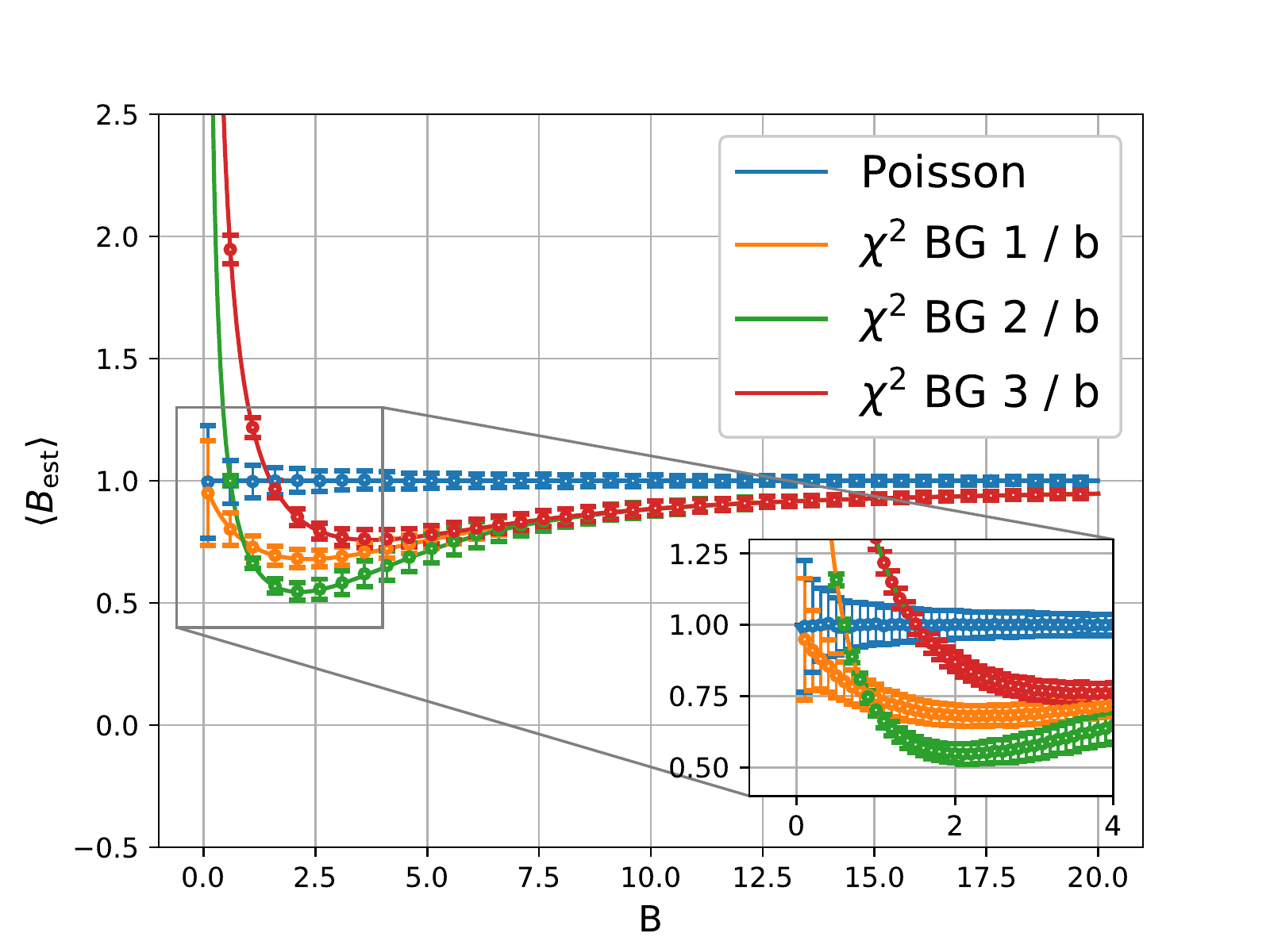}
    \caption{\textbf{Left}: Absolute and \textbf{Right}: relative background value minimizing the least squares method for the three cases in table~\ref{tab:ZeroCountTactic} denoted by points together with the analytical expressions in \ref{eq:LeastSquareMinimizing} signified by full curves. Inserts are zoom in on the low counts with all background values plotted while large figures contain every fifth point.}
    \label{fig:LeastSquaresBackgroundMinimizing}
\end{figure}

In the following the deviation of the background estimation using the least squares method on scattering data is investigated. Starting from the chi square
\begin{equation}
    \chi^2 = \sum_i\parh{\frac{n_i-\lambda_i}{\sigma_i}}^2,
\end{equation}
where $n_i$ is the count number in the $i$'th bin with $\sigma_i$ being the corresponding uncertainty estimate. $\lambda_i$ is the model prediction at $i$, which in the case of a flat background simply is $b$. To find the stationary point for this function, the first derivative with respect to the model parameter $b$ is found
\begin{equation}
    0 = \pdx[\chi^2]{b} = \sum_i-2\frac{n_i-b}{\sigma_i^2}
\end{equation}
Splitting the sums and isolating $b$ yields
\begin{equation}
    b = \parh{\sum_i\frac{1}{\sigma_i^2}}^{-1}\parh{\sum_i\frac{n_i}{\sigma_i^2}}.
\end{equation}
Next step is to split the two sums into a part containing the points that are zero and all the rest. It is here assumed that there are $z$ bins with zero counts and $m-z$ non-zero bins. This yields
\begin{equation}
    b = \parh{\sum_{i=1}^z\frac{1}{\tilde{\sigma}_i^2}+\sum_{i=z+1}^m\frac{1}{\sigma_i^2}}^{-1}\parh{\sum_{i=1}^z\frac{\tilde{n}_i}{\tilde{\sigma}_i^2}+\sum_{i=z+1}^n\frac{n_i}{\sigma_i^2}},\label{eq:bSums}
\end{equation}
where $\tilde{n}_i$ and $\tilde{\sigma}_i$ denote the values used for zero bins. These are different depending on the background strategy as presented in table.~\ref{tab:ZeroCountTactic}. First, the BG3 case is followed where zero counts  are removed and the uncertainty estimate is simply $\sigma_i = \sqrt{n_i}$. This results in
\begin{equation}
    b_3 = \parh{\sum_{i=z+1}^m\frac{1}{n_i}}^{-1}\parh{\sum_{i=z+1}^m\frac{n_i}{n_i}} = \frac{m-z}{\sum_{i=z+1}^m\frac{1}{n_i}},
\end{equation}
as the second sum merely is a sum of 1 with $m-z$ terms. If one assumes the number of bins measured is large, then the sum in the denominator can be approximated by the Poisson distribution of counts. That is the number of bins containing a certain number of counts $n$ is to distributed by the Poisson distribution given the expectation value, in this case $b$, multiplied with the total number of bins $m$, 
\begin{equation}
    \sum_{i=z+1}^{m} \frac{1}{n_i} \approx \sum_{n=1}^\infty P(n|b)m\frac{1}{n} = m\sum_{n=1}^\infty\frac{\mathrm{e}^{-b}b^n}{n!n}.
\end{equation}
In the above, the sum used in the approximation starts at 1 instead of 0 as the bins containing zero counts already have been taken care of. The sum can be performed yielding
\begin{equation}
    \sum_{n=1}^\infty \frac{\mathrm{e}^{-b} b^n}{n!n} = -\mathrm{e}^{-b} (\ln{-b} + \Gamma(0, -b) + \gamma ),
\end{equation}
where $\Gamma$ is the incomplete upper gamma function and $\gamma \approx 0.5772157...$ is the Euler constant. Then, the background estimate becomes
\begin{equation}
    b_3 = \frac{m-z}{m\parh{-\mathrm{e}^{-b} (\ln{-b} + \Gamma(0, -b) + \gamma )}} = \frac{1-\frac{z}{m}}{-\mathrm{e}^{-b} \parh{\ln{-b} + \Gamma(0, -b) + \gamma }},
\end{equation}
Following the same procedure for the two other strategies eq.~\ref{eq:bSums} becomes
\begin{align}
    b_1 &= \parh{z+\sum_{i=z+1}^m\frac{1}{n_i}}^{-1}\parh{0+\parh{m-z}} = \frac{1-\frac{z}{m}}{\frac{z}{m}+\parh{-\mathrm{e}^{-b} (\ln{-b} + \Gamma(0, -b) + \gamma )}} \\ 
    b_2 &= \parh{4z+\sum_{i=z+1}^m\frac{1}{n_i}}^{-1}\parh{2z+\parh{m-z}} = \frac{1+\frac{z}{m}}{4\frac{z}{m}+\parh{-\mathrm{e}^{-b} (\ln{-b} + \Gamma(0, -b) + \gamma )}}
\end{align}
To proceed, notice that the fraction $\frac{z}{m}$ is the number of zero count bins out the whole. This fraction is approximated by the probability of zero counts from the Poisson distribution, thus $\frac{z}{m} \approx P(0|b) = \mathrm{e}^{-b}$. Thus, the values minimizing the least squares value using the three background strategies are
\begin{subequations}\label{eq:LeastSquareMinimizing}
\renewcommand{\theequation}{\arabic{parentequation}b$_\arabic{equation}$}
\begin{align}
b_1 &= \frac{\mathrm{e}^{b}+1}{1-\parh{\ln{-b} + \Gamma(0, -b) + \gamma}} \label{eq:b1}\\
b_2 &= \frac{\mathrm{e}^{b}+1}{4-\parh{\ln{-b} + \Gamma(0, -b) + \gamma }}\label{eq:b2}\\
b_3 &= \frac{1-\mathrm{e}^{b}}{\ln{-b} + \Gamma(0, -b) + \gamma },\label{eq:b3}
\end{align}
\end{subequations}

and they are plotted in Fig.~\ref{fig:LeastSquaresBackgroundMinimizing} both with their absolute value and relative to the true background. If one stayed to the tactic of using $\sigma_i=\sqrt{n_i}$ independent of the value of $n$, it is immediately clear that the denominator in \ref{eq:bSums} is infinite due to the first term if a single zero count bin is present forcing the $b$ estimate to zero.

The limit for large background values is found for all three estimators in \eqref{eq:LeastSquareMinimizing} by identifying the important parts to be the exponential in the nominator and the incomplete upper gamma function in the denominator. The imaginary part from $\Gamma\parh{0,-b}$ is canceled by the imaginary part of $\ln{-b}$. In the large limit
\begin{equation}
    \Gamma\parh{0,b} \rightarrow_{b\rightarrow\infty} \mathrm{e}^{b}\parh{-\frac{1}{x}-\frac{1}{x^2}+\cdots},
\end{equation}
which then makes the fraction
\begin{equation}
    \frac{\mathrm{e}^{b}}{-\Gamma(0, -b)} \approx \frac{1}{\frac{1}{b}+\frac{1}{b^2}}.
\end{equation}
Looking at the deviation from the actual value one finds
\begin{equation}
    \frac{1}{\frac{1}{b}+\frac{1}{b^2}}-b = \frac{-b}{1+b} = \frac{-1}{1+\epsilon},
\end{equation}
where $\epsilon = \frac{1}{b}$ goes to 0 when $b$ goes to $\infty$. The limit is radially found from insertion
\begin{equation}
    \lim_{b\rightarrow\infty} b_i -b = \lim_{\epsilon\rightarrow0}\frac{-1}{1+\epsilon} = -1.
\end{equation}
It has thus been found that all of the three background strategies, $b_i$, in the limit of infinite background yields a value 1 too low as compared to the true Poisson mean.

\subsection{Sensitivity of Multinomial and Poisson fits}\label{sec:PoissonWrong}
\begin{figure}
    \centering
    \includegraphics[width=0.6\linewidth]{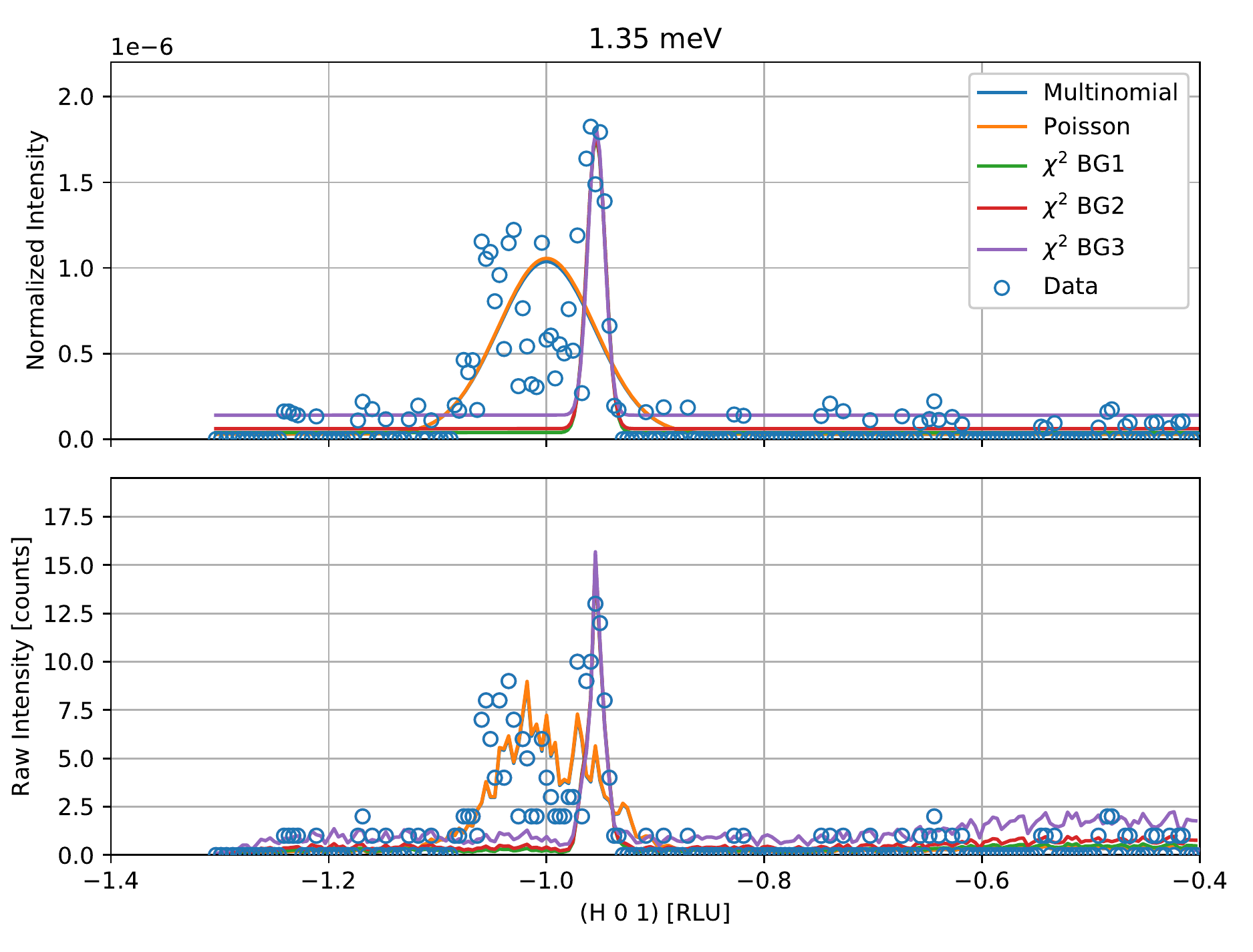}
    \caption{Fitting of MnF$_2$ signal at 1.35 meV where two distinct features are visible together with the fits using the Multinomial, Poisson and Gaussian approaches. Both the Multinomial and Poisson fits are seen to be heavily interfered by the presence of the second signal while the least squares method disregards it more easily.}
    \label{fig:WrongFit}
\end{figure}

During the fitting procedure of the MnF$_2$ it became apparent that the stability of the different methods were different, see Fig.~\ref{fig:WrongFit}. When performing the parameter estimations using the different techniques and the same initial guess only the Gaussian approach was robust against a model not fully capturing the data. Both the Multinomial and Poisson methods are influenced by the second feature in the data.  This is seen by the lowering of peak intensity and broadening of peak width balancing to fit both peaks. From the discussion of log-likelihoods shallowness of the Poisson and Multinomial log-likelihood as compared to the $\chi^2$ methods it is sensible the Poisson and Multinomial methods are less stable against data not explained by the model. Two ways of overcoming this exist; mask the data not explained by the model, i.e. $H<-1$, or extend the model to include both peaks. The latter suggestion introduces extra parameters to be fitted. However, in this particular case the position of the peaks are to be centered around -1 and their integrated intensities are to be equal. Extending this analysis requires a global model and, preferably, a model of the instrument effect on the data. With such an instrument model no excess parameters are introduced. 

\end{document}